\documentclass[manuscript, 14pt]{agujournal2019}
\usepackage{url} %this package should fix any errors with URLs in refs.
\usepackage{lineno}
\usepackage{amsmath}
\usepackage{natbib}
\usepackage{amssymb}
\usepackage[inline]{trackchanges} %for better track changes. finalnew option will compile document with changes incorporated.
\usepackage{soul}
\draftfalse

\journalname{Geophysical Research Letters}

\begin{document}

%%%%%%%%%%%%%%%%%%%%%%%%%%%%%%%%%%%%%%%%%%%%%%%
%  TITLE
%
% (A title should be specific, informative, and brief. Use
% abbreviations only if they are defined in the abstract. Titles that
% start with general keywords then specific terms are optimized in
% searches)
%
%%%%%%%%%%%%%%%%%%%%%%%%%%%%%%%%%%%%%%%%%%%%%%%

% Example: \title{This is a test title}

% UNCOMMENT
\title{Cosmogenic $^{10}$Be as a Tracer for Recent Heliospheric Encounters with Interstellar Cold Clouds and Nearby Supernovae}

%%%%%%%%%%%%%%%%%%%%%%%%%%%%%%%%%%%%%%%%%%%%%%%
%
%  AUTHORS AND AFFILIATIONS
%
%%%%%%%%%%%%%%%%%%%%%%%%%%%%%%%%%%%%%%%%%%%%%%%

% Authors are individuals who have significantly contributed to the
% research and preparation of the article. Group authors are allowed, if
% each author in the group is separately identified in an appendix.)

% List authors by first name or initial followed by last name and
% separated by commas. Use \affil{} to number affiliations, and
% \thanks{} for author notes.
% Additional author notes should be indicated with \thanks{} (for
% example, for current addresses).

% Example: \authors{A. B. Author\affil{1}\thanks{Current address, Antartica}, B. C. Author\affil{2,3}, and D. E.
% Author\affil{3,4}\thanks{Also funded by Monsanto.}}

% UNCOMMENT
\authors{
    Anna Nica\affil{1}, 
    Merav Opher\affil{1}, 
    Jesse A. Miller\affil{1}, 
    Jennifer L. Middleton\affil{2}, 
    {Joe Giacalone}\affil{3}
    % {Edward J. Brook\affil{4}}
}

\affiliation{1}{Department of Astronomy, Boston University, Boston, MA, USA}
\affiliation{2}{Lamont-Doherty Earth Observatory, Columbia University, Palisades, NY, USA}
\affiliation{3}{{Lunar and Planetary Laboratory, University of Arizona, Tucson, AZ, USA}}
% \affiliation{4}{{College of Earth, Ocean, and Atmospheric Sciences, Oregon State University, Corvallis, OR, USA}}

% Corresponding author mailing address and e-mail address:

% (include name and email addresses of the corresponding author.  More
% than one corresponding author is allowed in this LaTeX file and for
% publication; but only one corresponding author is allowed in our
% editorial system.)

% UNCOMMENT
\correspondingauthor{Anna Nica}{anica@bu.edu}

%%%%%%%%%%%%%%%%%%%%%%%%%%%%%%%%%%%%%%%%%%%%%%%
% KEY POINTS
%%%%%%%%%%%%%%%%%%%%%%%%%%%%%%%%%%%%%%%%%%%%%%%
%  List up to three key points (at least one is required)
%  Key Points summarize the main points and conclusions of the article
%  Each must be 140 characters or fewer with no special characters or punctuation and must be complete sentences

% Example:
% \begin{keypoints}
% \item	List up to three key points (at least one is required)
% \item	Key Points summarize the main points and conclusions of the article
% \item	Each must be 140 characters or fewer with no special characters or punctuation and must be complete sentences
% \end{keypoints}

\begin{keypoints}

\item {Production of $^{10}\text{Be}$ is sensitive to both heliospheric compression and interstellar cloud crossing duration}

\item {Weaker compressions can be detected only for crossings lasting $\gtrsim1$ Myr; stronger compressions can be detected for crossings lasting $\gtrsim10$ kyrs}

\item {The $^{10}\text{Be}$ peak observed by Koll et al. (2025) at 10 Ma cannot be attributed to a supernova, but could be attributed to an interstellar cloud}

\end{keypoints}

%%%%%%%%%%%%%%%%%%%%%%%%%%%%%%%%%%%%%%%%%%%%%%%
%
%  ABSTRACT and PLAIN LANGUAGE SUMMARY
%
% A good Abstract will begin with a short description of the problem
% being addressed, briefly describe the new data or analyses, then
% briefly states the main conclusion(s) and how they are supported and
% uncertainties.

% The Plain Language Summary should be written for a broad audience,
% including journalists and the science-interested public, that will not have 
% a background in your field.
%
% A Plain Language Summary is required in GRL, JGR: Planets, JGR: Biogeosciences,
% JGR: Oceans, G-Cubed, Reviews of Geophysics, and JAMES.
% see http://sharingscience.agu.org/creating-plain-language-summary/)
%
%%%%%%%%%%%%%%%%%%%%%%%%%%%%%%%%%%%%%%%%%%%%%%%

%% \begin{abstract} starts the second page

% UNCOMMENT
\begin{abstract}

%%%% within 150-word limit %%%%
Recent works suggest there are periods when the Sun encountered massive interstellar cold clouds which compressed the heliosphere to within Earth's orbit, exposing Earth to interstellar galactic cosmic rays and energetic particles of heliospheric origin. We model $^{10}\text{Be}$ production in Earth's atmosphere during possible interstellar cloud encounters and supernovae. {We find that, if the heliosphere is compressed to 0.2 AU ($r_{TS}=0.12$ AU), the $^{10}\text{Be}$ production rate is elevated $9\times$ above the background. To be distinguished in marine sediments (iron-manganese crusts) beyond terrestrial variability, this signal must be sustained for $>0.01$ Myr  ($>0.5$ Myr). A 0.7 AU ($r_{TS}=0.4$ AU) compression increases the $^{10}\text{Be}$ production rate $3\times$ above the background, which may be detectable if the event lasts $>1$ Myr.} We find that the $^{10}\text{Be}$ peak observed by Koll et al. (2025) 10 Ma {is too prolonged to be produced by} a supernova, but could be attributed to an interstellar cloud crossing.

\end{abstract}

% UNCOMMENT
\section*{Plain Language Summary}

%%%%% within 200 word limit %%%%%
{{The solar wind forms the heliosphere, which surrounds the Solar System as the Sun moves through interstellar space.}} The heliosphere can be compressed by dense, cold interstellar clouds, accelerating energetic particles towards Earth and exposing Earth to more galactic cosmic rays. This produces more of the radioisotope $^{10}\text{Be}$ in Earth's atmosphere, which is deposited in polar ice, deep-ocean sediments and iron-manganese crusts growing on the seafloor. We calculate the $^{10}\text{Be}$ signal during encounters with interstellar cold clouds. {We find that the $^{10}\text{Be}$ signal in Fe-Mn crusts, ocean sediments, and ice is sensitive to heliospheric compression and duration of the interstellar cloud crossings. Compressions to 0.7 AU produce a $3\times$ elevated signal and can be detected in Fe-Mn crusts if crossings last $\gtrsim1$ million years. Compressions to 0.2 AU produce a $9\times$ increase in the $^{10}\text{Be}$ signal, which can be detected in sediments and ice if they last at least 10,000 years, and in Fe-Mn crusts for crossings lasting at least 0.5 million years.} Our results suggest that the $^{10}\text{Be}$ peak 10 million years ago reported by Koll et al. (2025) {was too prolonged to have been caused by a supernova, but could have been caused by an interstellar cloud crossing.}

%%%%%%%%%%%%%%%%%%%%%%%%%%%%%%%%%%%%%%%%%%%%%%%
%
%  BODY TEXT
%
%%%%%%%%%%%%%%%%%%%%%%%%%%%%%%%%%%%%%%%%%%%%%%%

\section{Introduction}\label{sec:intro}

The Sun moves with a speed of 19 km/s relative to nearby stars, carrying the Solar System through diverse regions of the interstellar medium (ISM) with different densities, temperatures, and compositions. As the Sun travels, the solar wind engulfs the Solar System in a protective bubble known as the heliosphere. The heliosphere shields Earth from interstellar galactic cosmic rays (GCRs) that are accelerated by high-energy events occurring elsewhere in the galaxy. Voyager spacecraft observations have shown that the heliosphere deflects 80\% of GCRs with energies 70 MeV-5 GeV at its outer boundary \citep{cummings_galactic_2016, stone_cosmic_2019}. 

The shape and size of the heliosphere depends on the pressure balance between the outflowing solar wind and the surrounding ISM. Today, the heliopause extends to $\sim$120 AU in the Sun's direction of motion when the neutral H density of the ISM is $\sim$0.1-0.2 cm$^{-3}$ \citep{stone_voyager_2013, stone_cosmic_2019}. {The termination shock, where the solar wind flow transitions from supersonic to subsonic, sits at 84-94 AU in today's heliosphere \citep{stone_voyager_2005, richardson_cool_2008, stone_asymmetric_2008}.} However, this size could have changed dramatically if the Solar System passed through dense, cold interstellar clouds with densities $\geq900$ cm$^{-3}$. After tracing the trajectory of the Solar System back several million years, \cite{opher_passage_2024} and \cite{opher_possible_2024} predict that the Solar System may have traversed through interstellar cold clouds or regions 2-3 million years ago (Ma) and 6-7 Ma, respectively. {{These clouds are composed of dense, cold neutral hydrogen. Recent observations find that clouds may be as small as 200 AU \citep{vannier_mapping_2025, meyer_remarkable_2012}. However, using observations of supernova remnants (and edges of bubbles), it is reasonable to assume that the clouds could be parsec size \citep{cahlon_parsec_2024}.}}
%they may be as thin as 200 AU across and filamentary \citep{meyer_remarkable_2012, vannier_mapping_2025, peek_path-breaking_2025}.}}% A

{{In this work, we determine the detectability of potential crossings through interstellar clouds of different densities and sizes. We first consider a crossing through the Local Lynx Cold Clouds 2-3 Ma with density 3000 cm$^{-3}$ and a 20 km/s velocity relative to the motion of the Sun. This crossing compressed the termination shock to a distance of 0.12 AU from the Sun, based on the heliosphere simulation results of \cite{opher_possible_2024}. We then consider a crossing through the edge of the Local Bubble 6-7 Ma, with a density of 900 cm$^{-3}$ and relative velocity of 32 km/s. This compressed the termination shock to 0.4 AU, as seen in the heliospheric model results of \cite{opher_passage_2024}. Both of these encounters compress the heliosphere such that Earth spends a portion of its orbit inside the tail end of the heliosphere, and the other portion outside the heliosphere.} The timing of these encounters agrees with an increase of radioactive isotope $^{60}\text{Fe}$ found in deep sea sediments \citep{knie_indication_1999, knie_60Fe_2004, fitoussi_search_2008, ludwig_time_2016, wallner_recent_2016, wallner_60fe_2021}, lunar samples \citep{fimiani_interstellar_2016}, and in GCRs \citep{binns_observation_2016}. It is unclear whether the peaks in $^{60}\text{Fe}$ correspond to supernova explosions producing $^{60}\text{Fe}$ in those time periods, or whether supernovae seeded interstellar clouds with $^{60}\text{Fe}$, which the Solar System later passed through. We use these models as benchmarks to explore how varying heliospheric compression affects {energetic particle and cosmic ray} exposure on Earth, and how interstellar cloud and supernova signals may be distinguished in geologic records of $^{10}\text{Be}$ isotopes.}

{\cite{opher_increased_2026} found that heliospheric compression upon encountering a cold cloud exposes Earth to high fluxes of GCRs while Earth is outside the compressed heliosphere, and heliospheric energetic particles (HEPs) during periods when Earth is immersed in the tail side of the compressed heliosphere. HEPs can be accelerated to tens to hundreds of MeV at the compressed termination shock.} Earth's dipole magnetic field deflects particles with energies up to 15 GeV near the geomagnetic equator, however MeV-range GCRs and solar energetic particles can still enter Earth's atmosphere near the poles \citep{comedi_spatial_2020}. 
%{Upon entering the atmosphere, these particles undergo collisions with nitrogen and oxygen, and produce ion pairs and cosmogenic isotopes.}

Cosmogenic isotopes---including $^{10}\text{Be}$---are produced when cosmic rays and energetic particles collide with nuclei of atmospheric gases in Earth's atmosphere and induce nucleonic-muon-electromagnetic cascades \citep{poluianov_production_2016, miyake_extreme_2019, lal_cosmic_1967}. $^{10}\text{Be}$ has a half-life of 1.4 million years (Myr), which is long enough to detect radiation changes within the past 12 Myr \citep{willenbring_long-term_2010}. $^{10}\text{Be}$ produced in Earth's atmosphere is deposited on the ocean floor, and remains detectable thousands-to-millions of years later through sampling of marine sediments, deep-ocean iron-manganese (Fe-Mn) crusts, and ice cores. {Examples of such records are shown by \cite{beer_use_1990, raisbeck_evidence_2006, raisbeck_direct_2007, willenbring_long-term_2010}, and \cite{simon_increased_2018}}. 
Prolonged exposure to increased radiation should increase cosmogenic radionuclide production in Earth's atmosphere, and thus may be recorded in geologic archives of these isotopes, with the duration depending on the size of the interstellar cold cloud.

{We consider three types of geologic archives of $^{10}\text{Be}$ as candidates to detect an interstellar cloud crossing: ocean sediments, Fe-Mn crusts, and Antarctic ice cores.}

{{Ocean sediment cores are sampled from sedimentary layers which accumulate on the ocean floor. $^{10}\text{Be}$ is deposited within each sedimentary layer after transport via atmospheric and ocean circulation, allowing for changes in $^{10}\text{Be}$ production and deposition to be seen over time. Sediments typically accumulate at a rate scale of 1-10 cm/kyr, though this range can vary as a function of depositional environment. Each $^{10}\text{Be}$ measurement typically integrates 1 cm of material, yielding temporal resolutions as short as 1 kyr \citep{simon_increased_2018}. Ocean sediments typically see 2-4$\times$  variability in the $^{10}\text{Be}$ signal induced by terrestrial variations in production and deposition. These variations can include fluctuations in geomagnetic field intensity and climatic and environmental changes occurring on thousand to hundred-thousand-year time scales \citep{frank_200_1997, simon_increased_2018, middleton_oceanographic_2026}.}}

Fe-Mn crusts are mineral growths of iron and manganese oxides that precipitate on to the surface of a substrate rock or crust \citep{hein_iron_1997, hein_cobalt-rich_2000}. They accumulate much more slowly than ocean sediments at a rate of 1-5 mm/Myr, with each $^{10}\text{Be}$ measurement typically integrating 1 mm of material. This yields a 0.2-1 Myr time integration per sample. Integrating over these longer timescales, $^{10}\text{Be}$ data from Fe-Mn crusts typically exhibit terrestrial variability on the order of 1-2$\times$ \citep{willenbring_long-term_2010}.

Ice cores are core samples taken from glaciers and ice sheets. Cosmogenic isotopes accumulate in surface snow and ice via atmospheric deposition. {Continuous ice cores provide the highest-resolution geologic record of past $^{10}\text{Be}$ variability, but the oldest recovered continuous ice core only dates back to 0.8 Ma \citep{raisbeck_evidence_2006}. Discontinuous ice core samples can record changes in $^{10}\text{Be}$ without the complications of oceanographic processing and date as far back as the Miocene 6 Ma \citep{shackleton_miocene_2025}, though $^{10}\text{Be}$ has not yet been measured in these cores.} $^{10}\text{Be}$ signatures lasting $\sim$1 year due to solar proton events can be detected in ice cores {\citep{miyake_cosmic_2015, mekhaldi_multiradionuclide_2015, ohare_multiradionuclide_2019, paleari_cosmogenic_2022, bard_radiocarbon_2023}}. The $^{10}\text{Be}$ production rate measured in ice cores can change up to $\sim2\times$ due to changes in Earth's geomagnetic field intensity \citep{raisbeck_evidence_2006, raisbeck_direct_2007}.
% for typical variability: Raisbeck et al. 2006, delaygue et al. 2011
% Variations in $^{10}\text{Be}$ production as measured in ice cores have a 10\% uncertainty \citep{delaygue_antarctic_2011}

The detectability of a cold-cloud encounter using cosmogenic isotopes has been explored by sampling lunar soil and {Fe-Mn crusts}, though these methods are limited by the timing and duration of the crossing. \cite{poluianov_detectability_2025} model the $^{26}\text{Al}$ signal produced by an interstellar cloud encounter that would be detectable in lunar soil samples. $^{26}\text{Al}$ has a half-life of $\sim$0.7 Myr \citep{poluianov_solar_2018}, making it another cosmogenic isotope ideal for observing cold cloud crossings several million years ago. \cite{poluianov_detectability_2025} find no evidence that the heliosphere crossed through an interstellar cold cloud 2-4 Ma, but the authors suggest that cloud crossings shorter than a few tens of thousands of years would be undetectable with this approach. A cloud crossing event would need to last $\sim$100 kyrs to be detectable in existing lunar soil data, but shorter crossings could be detected using modern, higher-precision sampling techniques. It is not clear if the signal of the proposed cold cloud crossing between 2-3 Ma can be detected in records of $^{10}\text{Be}$, as the cloud size is currently unknown. 
{If the cloud size is on the order of AU for a relative speed of 20 km/s, the crossing time could be as short as a few hundred years.}
On the other hand, \cite{koll_cosmogenic_2025} detect a prolonged increase in $^{10}\text{Be}$ production 10 Ma lasting 1-2 Myrs and suggest an interstellar cold cloud crossing as a possible explanation. 
A 1-2 Myr crossing event could correspond to a cloud that is tens of parsecs across.

In this work, we use the Cosmic Ray Atmospheric Cascade model for atmospheric production of cosmogenic isotopes \citep{poluianov_production_2016} to predict the production rates of $^{10}\text{Be}$ on Earth during possible encounters with interstellar cold clouds. 
The detectability of an interstellar cloud crossing in the geologic record depends on the size, temperature, density, and relative velocity of the cloud. This work determines how increased radiation induced by an interstellar cold cloud could affect $^{10}\text{Be}$ production on Earth.
We explore several potential crossing times and evaluate whether such an event can be detected in terrestrial archives.

% \begin{figure}
%   \centering
%     \includegraphics[width=\textwidth]{heliosphere_diagram_abc.pdf}
%       \caption{
%       Cartoon diagram showing compression of the heliosphere and Earth's exposure to energetic particles as the Solar System intersects an interstellar cold cloud. Panel a shows heliospheric compression following a collision with a cold cloud in the ecliptic plane, from an edge-on and top-down view {{\citep{opher_possible_2024}}}. Earth's orbit dips in and out of the heliosphere, exposing Earth to interstellar GCRs while it is outside the heliosphere and to HEPs inside the heliosphere. {Panel b shows a larger heliosphere which is compressed at an inclined angle \citep{opher_passage_2024}}. Panel {c} shows heliospheric compression if the cloud's relative velocity is perpendicular to the ecliptic plane and compressed to within Earth's orbit.
%       }
%       \label{fig:heliosphere}
% \end{figure}

\begin{figure}
  \centering
    \includegraphics[width=0.7\textwidth]{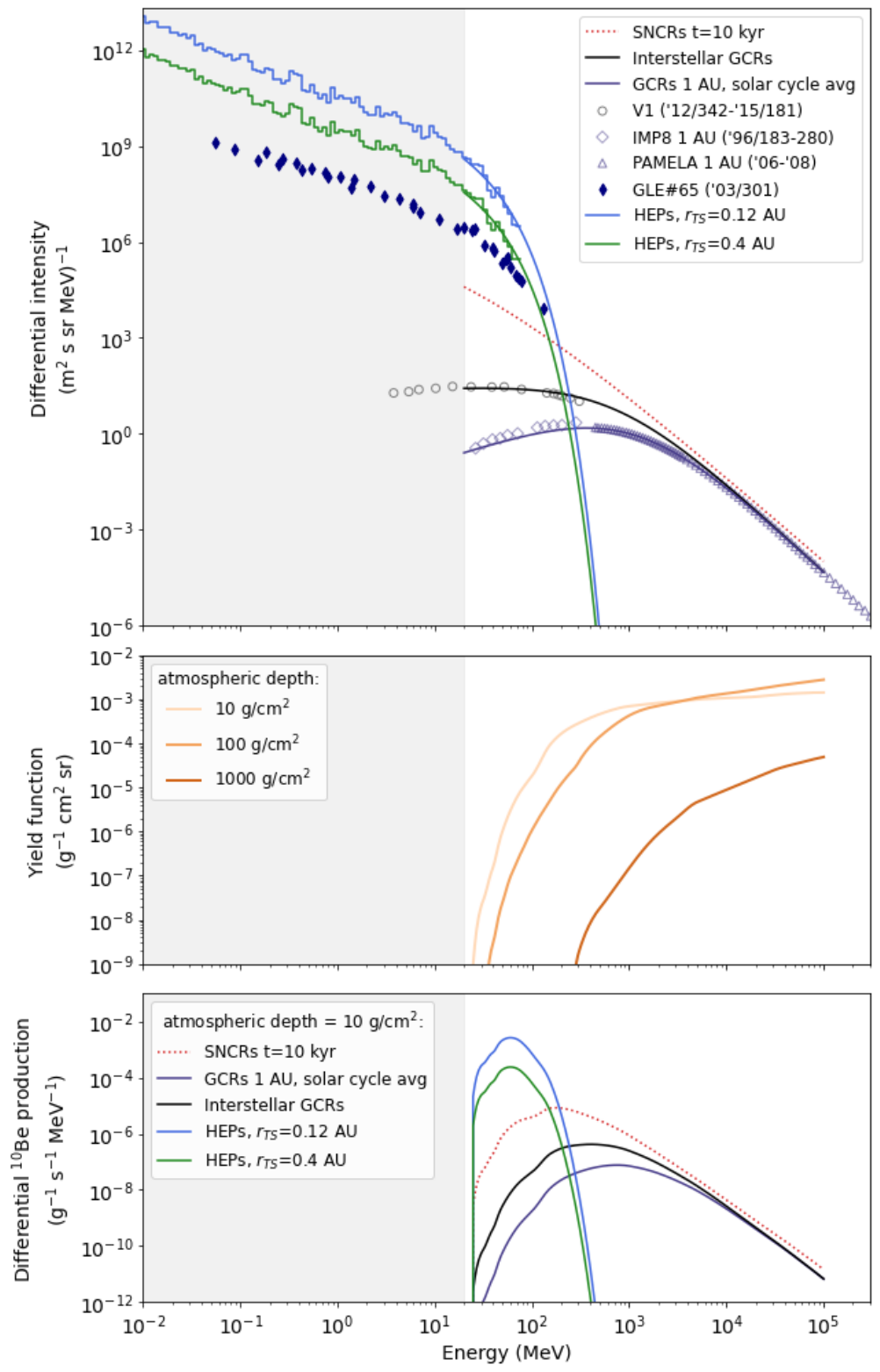}
      \caption{
      {(Top panel)} Proton differential energy spectra. The heliospheric energetic particle (HEP) flux \citep{opher_increased_2026} is shown for a termination shock distance of 0.12 AU (blue line), and 0.4 AU (green line), extended by Ellison-Ramaty fit functions to higher energies. {Gray shaded regions show the lower energy limit of the CRAC:10Be model.} Blue {diamonds} represent SEP fluxes during the 2003 Halloween solar storm \citep{mewaldt_proton_2005}. Black circles show Voyager 1 measurements of interstellar GCRs \citep{cummings_galactic_2016}. {The black line shows the interstellar GCR spectrum fit to these measurements \citep{vos_new_2015}.} The purple line shows the GCR flux reaching Earth in a modern-day heliosphere averaged across the solar cycle, and purple points represent GCR data from {PAMELA} and IMP8 at 1 AU {\citep{adriani_pamela_2011, mcdonald_cosmic_1998}}. The red dotted line shows the {peak} supernova cosmic ray (SNCR) spectrum {10 kyrs} after a supernova blast 50 pc away \citep{thomas_terrestrial_2023}.
      {(Middle panel) Yield functions for $^{10}\text{Be}$ shown at several atmospheric depths.}
      {(Bottom panel) Differential $^{10}\text{Be}$ production at an atmospheric depth of 10 g/cm$^2$ for the SNCR, GCR, and HEP spectra shown in the top panel.}
      }
      \label{fig:spectra}
\end{figure}

\section{Model Description} \label{sec:model} 
\subsection{Heliosphere and Radiation Model} \label{subsec:hel_rad_model}
Compression of the heliosphere during a cold cloud crossing would depend on the velocity of the cloud relative to the Sun {and the density of the cloud. Both parameters will influence the duration Earth will spend inside the compressed heliosphere, as well as the intensity of particles accelerated at the termination shock. The distance of the termination shock will largely influence that amount of radiation.} {In this work, we consider three possibilities. The first is a collision with a cold cloud in the ecliptic plane, where the termination shock distance of 0.12 AU from the Sun, {based on the heliospheric model of \cite{opher_possible_2024}}. The second is a collision where the termination shock is at 0.4 AU, based on the model of \cite{opher_passage_2024}. The third is a hypothetical collision perpendicular to the ecliptic plane.} In the first two cases, Earth would spend a portion of its orbit inside the heliosphere and a portion outside the heliosphere. In the perpendicular case, if Earth's orbit is sufficiently compressed to within Earth's orbit, Earth would be continuously exposed to interstellar space. 
%{Because the dimensions of the cold clouds are not known, we consider cloud crossing times of 1 kyr, 10 kyr, and 1 Myr}

%and a collision perpendicular to the ecliptic plane (Figure \ref{fig:heliosphere}). An ecliptic collision is an example of an event where Earth would spend a portion of its orbit inside the heliosphere and a portion outside the heliosphere (Figure \ref{fig:heliosphere}a). {This schematic is based on the heliosphere simulation results of \cite{opher_possible_2024} and \cite{opher_passage_2024}}. In a perpendicular collision (Figure \ref{fig:heliosphere}b), if the heliosphere is sufficiently compressed to within Earth's orbit, Earth would be continuously exposed to interstellar space. {Because the dimensions of the cold clouds are not known, we consider cloud crossing times between 100 years and 1 Myr. Assuming the relative speed between these interstellar clouds and the Sun is 20 km/s, a crossing time of 100 years means the Sun will have traversed $\sim$200 AU through the cold cloud, while a crossing time of 1 Myr suggests the Sun will have traversed through 1 pc of dense interstellar cloud material.}
% Because the dimensions of the cold clouds are not known, we consider cloud crossing times between 100 years and 1 Myr. This can correspond to clouds on hundreds-of-AU scales to tens-of-parsecs scales. Cloud sizes are based on a nominal speed between the cloud and the Sun of 20 km/s \citep{opher_possible_2024, opher_passage_2024}.

While outside the heliosphere, Earth experiences interstellar fluxes of GCRs. We assume minimal attenuation of GCRs by the cold cloud: a 1 GeV {proton-proton collision with cross-sectional area $4\times10^{-26}$ cm$^2$} has a mean free path of 2700 pc {in an interstellar cold cloud with density 3000 / cm$^3$ and a mean free path of 9000 pc in a cloud with density 900 / cm$^3$}, making collisions with protons within even a 20-pc cloud unlikely. \cite{morlino_cosmic_2015} find a $\sim$10\% reduction in flux for GCRs with energies 100 MeV within a cloud with a column density similar to the estimated size of the Local Leo Cold Cloud \citep{meyer_remarkable_2012}. Our resulting $^{10}\text{Be}$ production rates may differ by $\sim$10\% in the upper atmosphere, depending on the extent of the cloud. For the spectrum of interstellar GCRs, we use the parametrization by  \cite{vos_new_2015} fit to Voyager 1 measurements \citep{stone_voyager_2013}. Modulation of GCR fluxes reaching Earth throughout the solar cycle is parameterized using the modulation potential $\phi$ \citep{caballero-lopez_limitations_2004}. {\cite{poluianov_solar_2018} calculate the average solar modulation of GCRs over the past Myr to be 496 MV. For this study, we estimate an average $\phi=500$ MV.}

In {the first two} cases, Earth would re-enter the heliosphere for part of its orbit and be exposed to heliospheric energetic particles (HEPs) during this time \citep{opher_increased_2026}. HEPs are produced when energetic particles in the solar wind are accelerated at the termination shock of the compressed heliosphere. In a compressed heliosphere, the termination shock becomes a parallel shock. The termination shock is also much stronger (shock compression of 4) and accelerates high fluxes of energetic particles from keV to MeV energies \citep{opher_increased_2026}. As shown in Figure \ref{fig:spectra}, the flux of HEPs at MeV energies is at least $\sim$5-6 orders of magnitude greater than the flux of interstellar GCRs. The HEP flux is 1-2 orders of magnitude greater than the flux from a very large SEP event associated with the 2003 Halloween storm period \citep{mewaldt_proton_2005}. While effects from solar storms can last days to weeks, Earth's exposure to HEPs would occur for several months each year during an interstellar cloud crossing.

The HEP spectrum is based on a 1D model that solves the Parker transport equation at the termination shock of a compressed heliosphere \citep{opher_increased_2026}. {{We assume HEPs maintain constant intensities as they propagate beyond the termination shock, and thus uniformly fill the heliosheath. In the Parker solution, particle intensities remain constant downstream a shock. This is consistent with Voyager 1 observations, which saw an increase in ion intensities before the termination shock, then approximately constant intensities through the heliosheath \citep{decker_mediation_2008, decker_no_2012, krimigis_energetic_2019}.} Energetic particle flux at the termination shock scales with the density of the solar wind; hence, it falls with $1/r^2$. The HEP spectrum from \cite{opher_increased_2026} was calculated for a termination shock distance of 0.12 AU, so we scale the HEP flux for a heliosphere with a different termination shock distance proportional to:}
\begin{equation}\label{eq:TS}
    J_{HEP}(E)\propto\left(0.12\text{ AU}/r_{TS}\right)^2.
\end{equation}
where $J_{HEP}(E)$ is the HEP flux per energy $E$ at a given termination shock distance $r_{TS}$ from the Sun. {We apply an Ellison-Ramaty fit \citep{ellison_shock_1985}, commonly applied to SEPs, to the simulated HEP spectrum. SEPs and anomalous cosmic rays tend to have ratios of 20 protons to alpha particles; we apply the same ratio to HEPs \citep{mewaldt_proton_2005, giacalone_anomalous_2022}.}
% {As a lower limit for HEP-induced $^{10}\text{Be}$ production, we do not apply another fit function to the HEP spectrum, and assume all fluxes beyond the model's upper limit of 80 MeV are zero. We find that implementing an Ellison-Ramaty spectral fit \citep{ellison_shock_1985}, commonly applied to SEPs, yields a difference in yearly-average $^{10}\text{Be}$ production of $<1$\% from no fit.}

{The supernova cosmic ray (SNCR) {proton} spectrum is determined based on the model proposed by \cite{thomas_terrestrial_2023} for a supernova 50 pc away. There is a delay in the peak SNCR flux reaching Earth due to the travel time of SNCRs across 50 pc: SNCR flux peaks $\sim$10 kyrs after the supernova, then decreases over time across all energies.}

\subsection{$^{10}\text{Be}$ Production Model} \label{subsec:10Be_prod_model}
We compute the production of $^{10}$Be using the CRAC (Cosmic Ray Atmospheric Cascade) model developed by \cite{poluianov_production_2016} and validated by \cite{golubenko_application_2021}. This model numerically calculates yield functions of $^{10}\text{Be}$ using the GEANT4 simulation \citep{allison_geant4_2006, agostinelli_geant4simulation_2003} for primary protons and $\alpha$-particles with energies 20 MeV-100 GeV.  The model has an altitude range from 0-35 km, with an integrated bin for the upper atmosphere set at 100 km. 
% \textcolor{gray}{The production rate $Q$ is defined as}
% \begin{equation}\label{Q_eq}
%     \textcolor{gray}{Q(t,h,R_c)=\sum_i\int^\infty_{E_{c,i}}Y_i(E,h)\cdot J_i(E,t)\cdot dE,}
% \end{equation}
% \textcolor{gray}{where the yield function $Y_i(E,h)$ for each species $i$ depends on energy $E$ and atmospheric depth $h$, and the energy spectrum $J_i(E,t)$ depends on particle energy and time. Integration over energy begins above the energy cutoff $E_{c,i}$ for each species, which depends on geomagnetic field parameters.} 
We apply a geomagnetic dipole moment $M=7.8\cdot10^{22}$ Am$^2$, which is consistent with {time-averaged} paleointensity data from the present day to the Plio-Pleistocene era \citep{ biggin_paleointensity_2010}.

\subsection{Model Setup} \label{subsec:experiment}
We determine cosmogenic $^{10}\text{Be}$ production for cases when Earth was exposed to radiation due to cold cloud crossings and due to a nearby supernova (50 pc away). We use GCR spectra for the average solar modulation at Earth {over the past Myr of solar cycles ($\phi=500$ MV) \citep{poluianov_solar_2018}}, and interstellar GCRs while Earth is outside the compressed heliosphere ($\phi=0$ MV). {During cold cloud crossings, we explore cases when heliospheric compression exposed Earth to HEPs. We consider two cases: one with a termination shock at 0.12 AU, and one with a termination shock at 0.4 AU, scaled according to Equation \ref{eq:TS}. We estimate that Earth will be exposed to HEPs for 20\% of its orbit if the termination shock is at 0.12 AU {\citep{opher_possible_2024}}, and for 50\% of its orbit if the termination shock is at 0.4 AU \citep{opher_passage_2024}. We compare this to {cosmic ray} exposure from a supernova 50 pc away \citep{thomas_terrestrial_2023}.} We hold the geomagnetic field strength constant at present-day values. We do not account for atmospheric or ocean mixing within the model setup. Because the cloud sizes are not known, we consider cloud crossing periods lasting {10 kyrs, 100 kyrs, and 1 Myr. {At a relative speed of 20 km/s, these crossing times correspond to cloud sizes of 0.2, 2, and 20 pc, respectively.}} {We model the $^{10}\text{Be}$ signal that could be measured in Fe-Mn crusts, ocean sediments, and ice cores. We apply time integrations of 0.3 Myr for Fe-Mn crusts, and 10 kyr for ocean sediments and ice cores. Higher-resolution sampling of ocean sediments and ice cores is possible but resource-intensive for a 2-7 Ma time period. Since the exact timing of the interstellar cloud crossings is not known, we prioritize the ability to sample a larger time frame.}
% \citep{heikkila_meridional_2009, golubenko_full_2024, zheng_modeling_2024}.

% We determine cosmogenic $^{10}\text{Be}$ production for .....x.... cases: average solar modulation across a typical solar cycle ($\phi=500$ MV), exposure to interstellar GCRs while Earth is outside the compressed heliosphere ($\phi=0$ MV), and exposure to HEPs while Earth re-enters the compressed heliosphere. {For the compressed heliosphere, we explore two cases: one with a termination shock at 0.12 AU \citep{opher_possible_2024}, and one with a termination shock at 0.4 AU \citep{opher_passage_2024}.} We hold the geomagnetic field strength constant at present-day values. We consider cloud crossing periods lasting {1 kyr, 10 kyr, and 1 Myr.} We determine $^{10}\text{Be}$ production during an event where Earth enters and exits the heliosphere, estimating that Earth is exposed to interstellar GCRs for 80\% of its orbit and to HEPs for 20\% of its orbit \citep{opher_increased_2026}. These proportions match the heliospheric model presented by \cite{opher_possible_2024}, though exposure to HEPs and GCRs can change depending on the direction the Sun encounters a cold cloud and the shape of the heliosphere's tail. We repeat this process for a perpendicular collision, where Earth is exposed only to interstellar GCRs for its entire orbit.  We do not account for atmospheric mixing \citep{heikkila_meridional_2009, zheng_modeling_2024}.

\section{Results} \label{sec:results}
The density and relative velocity of an interstellar cloud, and therefore the amount of heliospheric compression and HEP exposure, can dramatically affect $^{10}\text{Be}$ production. A heliosphere compressed to {a heliopause distance of 0.2 AU} (such that $r_{TS}=0.12$ AU), produces {$\sim9\times$} greater yearly-averaged signal than {the background rate from modulated GCRs reaching Earth in a non-compressed heliosphere}. A heliosphere compressed to {0.7 AU} ($r_{TS}=0.4$ AU) produces {3$\times$} greater yearly-averaged signal. The $^{10}\text{Be}$ production rate from interstellar GCRs alone increases by $<$3$\times$. Earth’s dipole magnetosphere allows low-energy cosmic rays and energetic particles to enter the lower atmosphere at polar latitudes, but prevents particles with energies $<1$ GeV from entering the lower atmosphere below $\sim60^{\text{o}}$ latitude \citep{comedi_spatial_2020}. {{In Earth's atmosphere, $^{10}\text{Be}$ is typically produced with a threshold energy $\sim40$ MeV}, and HEP fluxes fall well below typical GCR fluxes at energies $>1$ GeV. Since HEPs only produce $^{10}\text{Be}$ in the MeV energy range, most of this $^{10}\text{Be}$ production occurs at polar latitudes, similarly to SEPs \citep{golubenko_full_2024}.}

{{Due to large-scale atmospheric dynamics, $^{10}\text{Be}$ deposition in marine sediments, Antarctic ice, and Fe-Mn crusts during the cloud crossing time period may not follow the same spatial pattern as $^{10}\text{Be}$ production in the atmosphere. Atmospheric models suggest that 50-80\% of $^{10}\text{Be}$ produced at latitudes 60-90$^\text{o}$N is deposited at latitudes lower than 60$^\text{o}$N \citep{heikkila_meridional_2009, zheng_modeling_2024}. Therefore, 50-80\% of $^{10}\text{Be}$ produced at polar latitudes in the model may instead be deposited at the surface at mid-latitudes, while only 20-50\% of the $^{10}\text{Be}$ produced in the polar atmosphere will ultimately be deposited at the poles. For sediments and Fe-Mn crusts, this spatial pattern can be further muddied by ocean circulation and dynamic sedimentary environments \citep{savranskaia_disentangling_2021, savranskaia_removing_2024, middleton_oceanographic_2026}.}}
% The presence of HEPs dramatically increases $^{10}\text{Be}$ production rates. HEP-induced polar $^{10}\text{Be}$ production increases by a factor of 1450 compared to the average point of the solar cycle. The HEP-induced globally-averaged $^{10}\text{Be}$ production increases by a factor of 270 from the average point of a modern-day solar cycle. Because we do not include atmospheric mixing, these production values over-predict the polar HEP-induced $^{10}\text{Be}$ deposition. Atmospheric models suggest that 50-80\% of $^{10}$Be produced at latitudes 60-90$^\text{o}$N is deposited at latitudes lower than 60$^\text{o}$N; conversely, $^{10}$Be produced at lower latitudes can be deposited near the poles \citep{heikkila_meridional_2009, zheng_modeling_2024}. Since HEPs are confined to MeV energies and lower, they do not penetrate geomagnetic shielding at equatorial latitudes, so HEP-induced $^{10}\text{Be}$ production occurs only near the poles. Measurements taken at ice core sites in Greenland or Antarctica may therefore see HEP-induced $^{10}\text{Be}$ deposition rates that are 20-50\% lower than the polar production rates we compute. 

\begin{figure}
  \centering
    \includegraphics[width=1.0\textwidth]{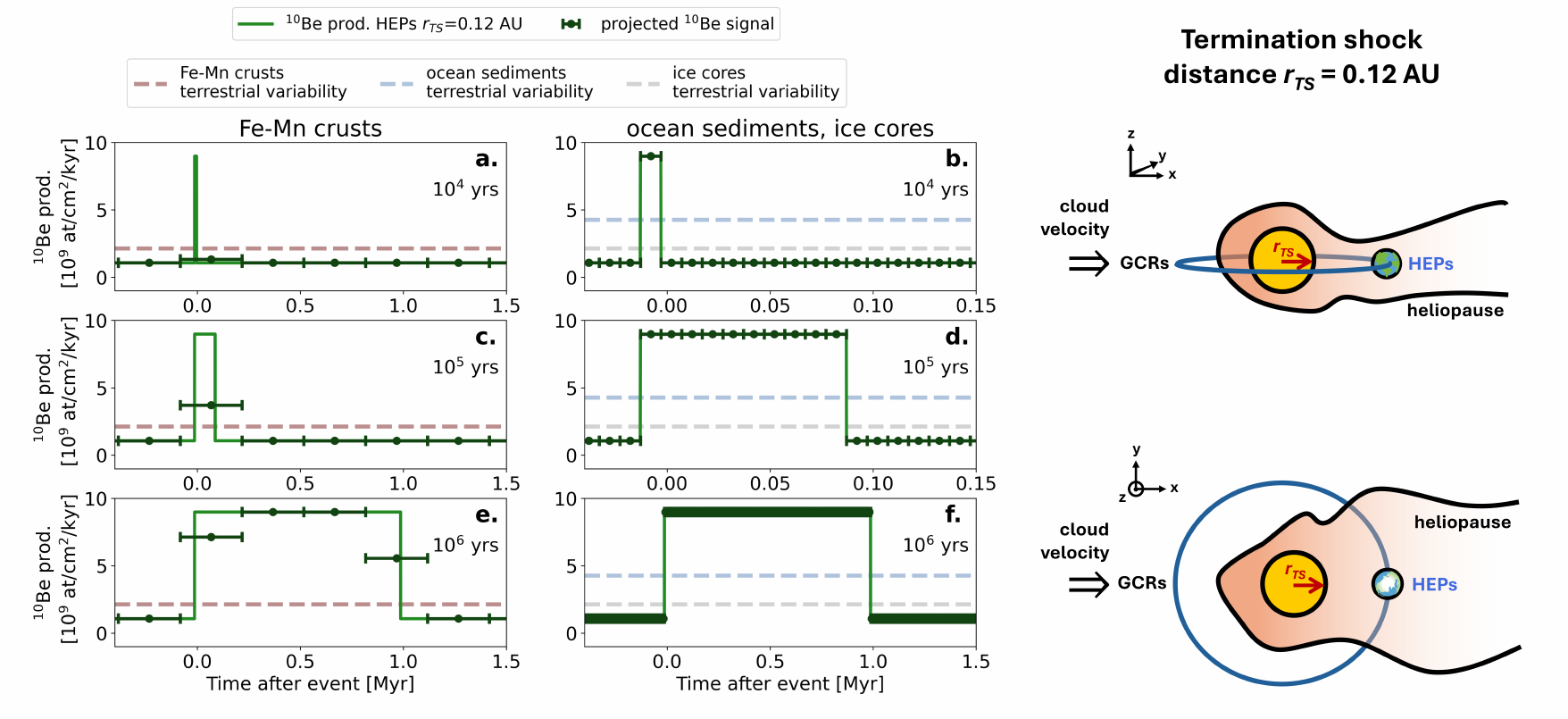}
      \caption{
      Simulated $^{10}\text{Be}$ signal during an ecliptic collision with an interstellar cloud {which compresses the heliosphere to a termination shock distance $r_{TS}=0.12$ AU.} We consider cloud crossing times of $10^4$, $10^5$, and $10^6$ years. Green lines show the modeled $^{10}\text{Be}$ production rate over time from 80\% GCRs and 20\% HEPs. Dark green points in the left panels show the projected signal in Fe-Mn crusts at 0.3 Myr resolution; points in the right panels show the projected signal in ocean sediments {and ice cores at 10 kyr resolution}. {Red, blue, and gray} dashed lines show upper limit of variability {within Fe-Mn crusts, ocean sediments, and ice core archives, respectively,} due to dynamic terrestrial processes  \citep{willenbring_long-term_2010, delaygue_antarctic_2011, simon_increased_2018, middleton_oceanographic_2026}. {On the right, a cartoon diagram shows compression of the heliosphere with $r_{TS}=0.12$ AU from an edge-on and top-down view {\citep{opher_possible_2024}}.}
      }
      \label{fig:10Be_time_avg}
\end{figure}

\begin{figure}
  \centering
    \includegraphics[width=1.0\textwidth]{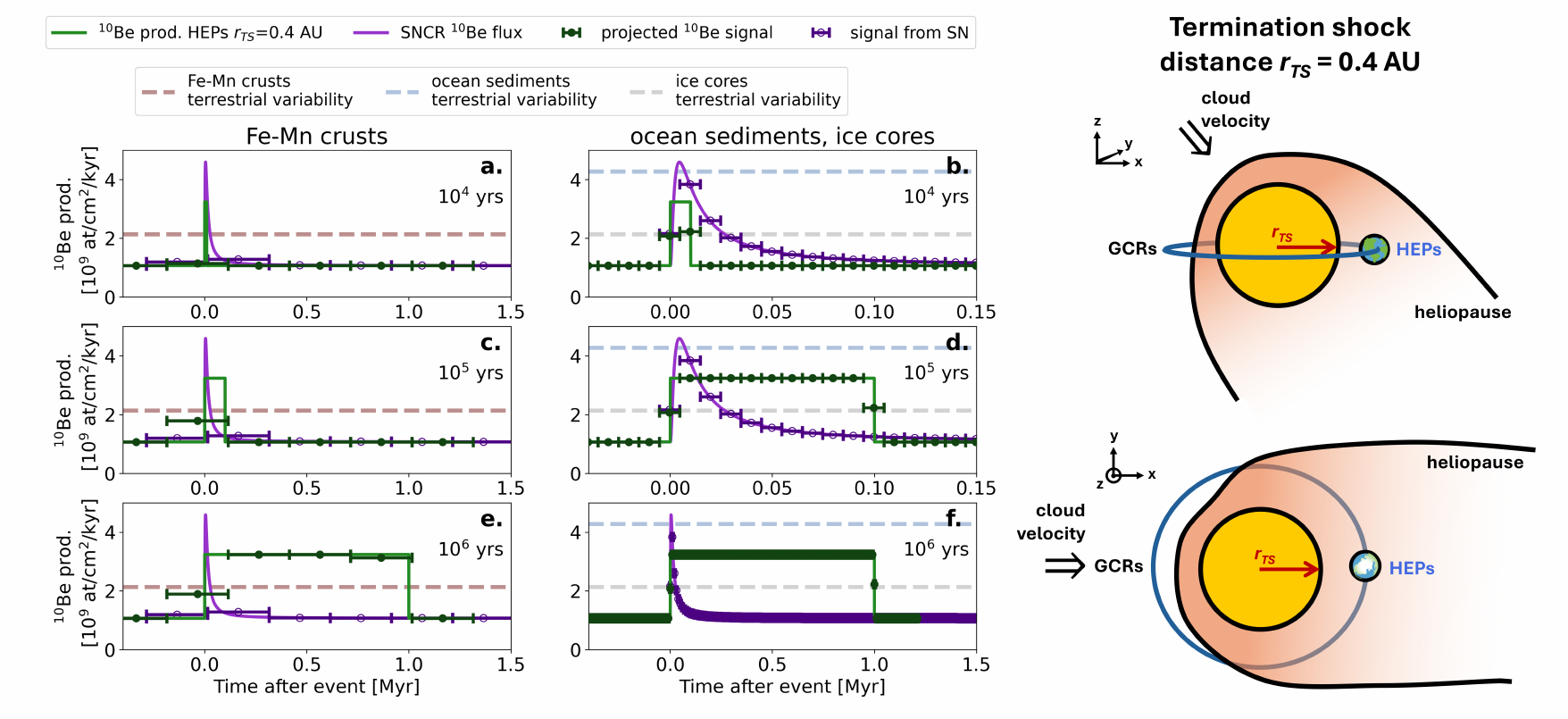}
      \caption{
      Similar to Figure \ref{fig:10Be_time_avg}, but for a less-compressed heliosphere that has a termination shock distance $r_{TS}=0.4$ AU. Green lines show the $^{10}\text{Be}$ production rate from 50\% GCRs and 50\% HEPs, and dark green points show the projected signal. Purple curves show the $^{10}\text{Be}$ production rate from a supernova 50 pc away, and dark purple points show the projected signal from this event. {On the right, a cartoon diagram shows compression of the heliosphere with $r_{TS}=0.4$ AU \citep{opher_passage_2024}.}
      }
      \label{fig:10Be_time_HEPs_SN}
\end{figure}

\begin{figure}
  \centering
    \includegraphics[width=1.0\textwidth]{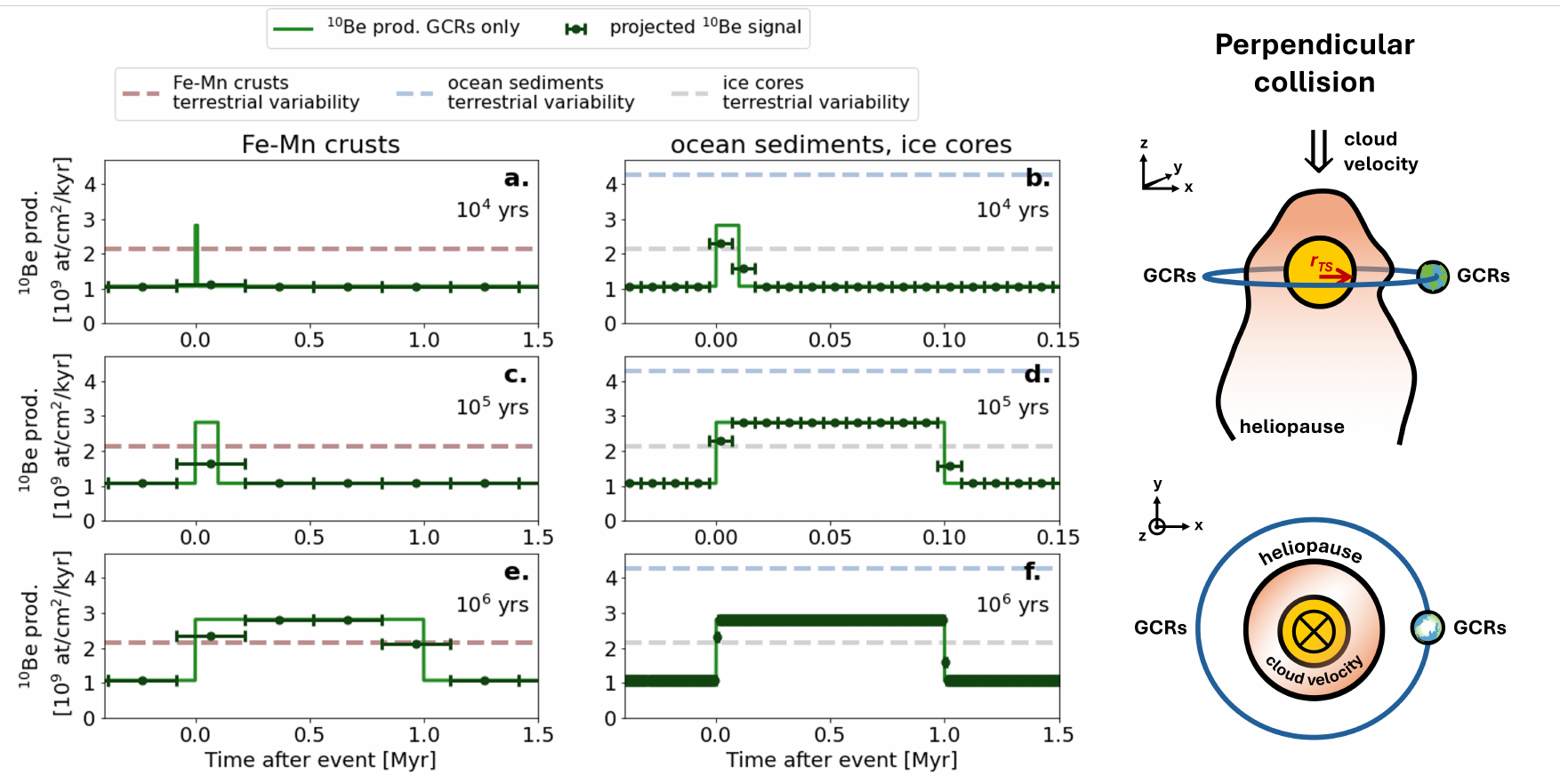}
      \caption{
      Similar to Figures \ref{fig:10Be_time_avg} and \ref{fig:10Be_time_HEPs_SN}, but for a perpendicular collision with a cold cloud. During the crossing, the $^{10}\text{Be}$ production rate is determined only by the interstellar flux of GCRs. {On the right is a cartoon diagram of a perpendicular compression of the heliosphere, where cloud velocity points into the page in the top-down view.}
      }
      \label{fig:10Be_time_GCR}
\end{figure}

While a cold-cloud encounter could increase $^{10}\text{Be}$ production in Earth's atmosphere, its detectability depends on {heliospheric compression}, cloud size, terrestrial variability in $^{10}$Be production and deposition, and the rate at which Earth was exposed to HEPs and GCRs. In Figure \ref{fig:10Be_time_avg}, we show a time series of the $^{10}\text{Be}$ production rate while Earth {orbited through a compressed heliosphere with $r_{TS}=0.12$ AU}. We model a cloud crossing that may have occurred 2-3 Ma, though our results can be generalized to other cloud-crossing events. Outside the cloud, we calculate $^{10}\text{Be}$ production induced by GCRs at the average modulation of the solar cycle, assuming the heliosphere and ISM had similar size and density to today. During the crossing, we assume a ratio of 80\% interstellar GCR exposure to 20\% HEP exposure per year on Earth {\citep{opher_possible_2024}.} This corresponds to HEP exposure for over 2 months each year. 
We then calculate the time-integrated $^{10}$Be signal that would be recorded in marine archives using typical time resolutions for ocean sediments and Fe-Mn crusts. Ice or seafloor deposits of $^{10}\text{Be}$ can be {represented as $^{10}\text{Be}$/$^9\text{Be}$ ratios or as $^{10}\text{Be}$ fluxes calculated from measured concentrations and material accumulation rates,} from which changes to the $^{10}\text{Be}$ production rate are inferred. 

We find that $^{10}\text{Be}$ can increase by a factor of {9} due to the high HEP flux. A crossing lasting 10 kyrs can be detected in ocean sediments and ice cores with time resolutions of 10 kyrs. A crossing must last longer than 10 kyrs, on the order of $\geq$0.1 Myr, to be detected in Fe-Mn crusts. {Due to the long integration time per sample of Fe-Mn crusts, a crossing lasting 0.1 Myr would measure as only a $4\times$ increase in $^{10}\text{Be}$ above the background, while a crossing lasting 1 Myr would measure as a $9\times$ increase above the background.} 
%This would produce a signal in a few mm of Fe-Mn crust (or $\sim$4 samples), which limits detailed characterization.}

% A short crossing on the order of 100 years ($\sim$400 AU cloud extension at a 20 km/s relative velocity) may be detectable beyond typical variability in ocean sediments, but is not detectable in Fe-Mn crusts (Figures \ref{fig:10Be_time_avg}a, \ref{fig:10Be_time_avg}b). If the crossing lasts on the order of 100 kyr (0.1 Myr, $\sim$2 pc cloud extension), it would produce an elevated $^{10}\text{Be}$ production signal spanning $\sim$1 m of sediment (or 100 samples at a 1 cm sampling resolution). A high density of measurements across a peak can allow for more detailed characterization of the nature of the crossing and the cold cloud's structure. In comparison, this scenario would produce an elevated $^{10}\text{Be}$ production signal in only a few mm of Fe-Mn crust (or 1-2 samples at a 1 mm sampling resolution), which limits any detailed characterization. A crossing lasting on the order of 1 Myr ($\sim$20 pc cloud extension) should enable multiple detections from both types of archives, at typical sampling resolution, if the HEP-induced signal is as high as 70$\times$.

In Figure \ref{fig:10Be_time_HEPs_SN}, we compare the $^{10}\text{Be}$ signal from a compressed heliosphere with $r_{TS}=0.4$ AU to a potential supernova signal occurring at the same time. These scenarios serve as alternate explanations for the observed peaks in $^{60}\text{Fe}$. For the interstellar cloud crossing, we use a ratio of 50\% HEPs and 50\% interstellar GCRs \citep{opher_passage_2024}. The cold cloud crossing and supernova events have similar $^{10}\text{Be}$ signals {3-4$\times$} the baseline. We find that a supernova is too short to be detectable in Fe-Mn crusts, but a cold cloud which compresses the {termination shock} to 0.4 AU can produce a distinguishable signal if the crossing lasts on the order of {1 Myr}. {Due to a wider range of terrestrial variability captured in the higher-resolution marine sediment records relative to Fe-Mn crusts, a short-duration 3-4$\times$ increase in $^{10}\text{Be}$ production would be more difficult to identify in a single sediment record. However, marine sediments could be used to resolve the difference between a supernova and an interstellar cold cloud crossing for crossing durations $>0.1$ Myr, if multiple sediment records spanning the same interval are examined and oceanographic sources of $^{10}\text{Be}$ variability can be constrained over the interval of interest (e.g., \citealp{frank_200_1997, savranskaia_removing_2024}).} When observed using ice core records, a supernova event would produce a distinct profile, with a sharp initial peak and drop-off lasting $\sim$75 kyr. An interstellar cold cloud, on the other hand, may produce either a distinct peak in the event of a {10 kyr crossing}, or a consistent signal {above the range of terrestrial variability} over the course of a longer crossing.

In Figure \ref{fig:10Be_time_GCR}, we show the $^{10}\text{Be}$ production rate over time for a perpendicular collision where the heliosphere may be compressed completely within Earth's orbit, exposing Earth only to interstellar GCRs. {This case serves as a baseline to compare to the effects of HEPs.} $^{10}\text{Be}$ production increases by less than a factor of 3 during the cloud crossing. This is completely within the scope of typical terrestrial variability for ocean sediments, making such a crossing signal harder to identify. We find that a cloud crossing cannot be detectable in Fe-Mn crusts if the event lasts $\leq$0.1 Myr because the $^{10}\text{Be}$ signal would be smoothed out within the integration time of a single Fe-Mn 1 mm sample. A cloud crossing where Earth is only exposed to interstellar GCRs could be detectable if the event lasts close to 1 Myr. However, this signal may be more difficult to distinguish from terrestrial causes that could also produce a long, low peak. Due to their high-latitude locations and independence from ocean processes, ice cores are better suited to detect external $^{10}\text{Be}$ production rate changes relative to {background changes in $^{10}\text{Be}$ deposition associated with variable terrestrial processes, provided sufficient ice core sample material is available over the interval of interest.}

\section{Discussion} \label{sec:discussion}
Comparing the modeled signal in Figures \ref{fig:10Be_time_avg}, {\ref{fig:10Be_time_HEPs_SN},} and \ref{fig:10Be_time_GCR}, it is evident that prolonged and periodic exposure to HEPs is necessary for interstellar cloud encounters lasting less than $\sim$1 Myr to be detectable in geologic archives. {While there are no continuous ice core records spanning $>0.8$ Ma, or existing high-resolution $^{10}\text{Be}$ records from ocean sediments spanning 6-7 Ma,} a crossing as short as 5-10 kyrs {which exposes Earth to HEPs} could be resolved using {even discrete} ocean sediment {or ice core} records if the timing of the event can be constrained to narrow the sampling period. A cloud crossing lasting $\leq\sim$10 kyrs cannot be detected in Fe-Mn crusts with 0.3-Myr time integrations. {Since cloud crossings have not been detected in the 2-3 Ma and 6-7 Ma time periods, the timing of smaller interstellar cloud crossings may need to be constrained astrophysically before selecting a sampling period for high-resolution records.}

% Comparing the {$^{10}\text{Be}$ production rate and projected signal} in Figures \ref{fig:10Be_time_avg}, {\ref{fig:10Be_time_HEPs_SN},} and \ref{fig:10Be_time_GCR}, we find that detectability of a cold-cloud encounter depends on the density and extension of the cloud. If the heliosphere crossed through a very thin or filamentary cloud, a crossing event may have occurred on a timescale of {$\sim$10 kyr} and would not be detectable in Fe-Mn crust records. % this is pretty much the same as above ...
%We find that high-resolution archives, such as those available in ocean sediments {and ice cores}, are necessary to detect such a short-lived encounter. 
%Prolonged and periodic exposure to HEPs is necessary for cold-cloud encounters lasting less than $\sim$1 Myr to be detectable in ocean sediments {and ice cores}. 
% An ecliptic collision lasting as short as 5-10 kyrs could be resolved using ocean sediments {or ice cores} if the timing of the event can be constrained to narrow the sampling period.
%Long-lived encounters, on the other hand, may have their detectability compromised if the cold cloud has a high enough density to attenuate GCRs to within typical variability. 

As discussed in Section \ref{sec:intro}, \cite{poluianov_detectability_2025} find that lunar samples of $^{26}\text{Al}$ can only detect crossing events lasting $\sim$100 kyr. At a relative velocity of 20 km/s, this would require crossing through a $\sim$2 pc cold cloud. Similarly, we find that a crossing must last on the order of 100 kyrs-1 Myr to be detected using Fe-Mn crust samples of $^{10}\text{Be}.$
\cite{koll_cosmogenic_2025, koll_no_2026} do not see anomalous increases of $^{10}\text{Be}$ in Fe-Mn crusts 2-3 Ma or 6-7 Ma, as proposed by \cite{opher_possible_2024} and \cite{opher_passage_2024}. {\cite{opher_reply_2025} argue that this is due to the coarse time resolution of Fe-Mn crusts and limited resolution of available sediment data, such that small crossings will not be detected in such samples. Recent}{ sediment data do not show prolonged changes in $^{10}\text{Be}$ (relative to terrestrial variations in the geomagnetic field) from 0-4 Ma either \citep{valet_geomagnetic_2025}.} {Our results suggest that there are scenarios where an interstellar cloud crossing may have occurred, but would be undetectable in terrestrial geologic records, such as those shown by \cite{koll_no_2026}: for example, a crossing lasting 10 kyrs or shorter where Earth is only exposed to interstellar GCRs.}
% {\cite{koll_no_2026} suggest that, if a cloud crossing occurred 2-3 Ma and exposed Earth to increased GCR intensity, the crossing would need to have lasted on the order of $100$ kyrs or longer to be detected in Fe-Mn crusts. Our results suggest that a crossing lasting on the order of $10$ kyrs could be detected in Fe-Mn crusts if the compressed heliosphere is oriented such that HEPs reach Earth. } 

\cite{koll_cosmogenic_2025} report a prolonged increase in $^{10}\text{Be}$ concentrations in Fe-Mn crusts around 9-11 Ma that could be attributed to an interstellar cold cloud crossing. These measurements are elevated beyond typical variability for $\sim$1-2 Myr, and the peak detection is a factor of 1.7 greater than background measurements. {A peak with a Myr-scale duration cannot be attributed to a single supernova explosion at that time, since the increase in $^{10}\text{Be}$ from a supernova would last $<$0.1 Myrs.} If the peak seen by Koll et al. can be attributed to an increase in flux during a cold cloud crossing, it would be consistent with our interstellar GCRs case (Figure \ref{fig:10Be_time_GCR}f), {or a case with less heliospheric compression {(Figure \ref{fig:10Be_time_HEPs_SN}f)}.} If a cloud crossing occurred during this time, it may have compressed the heliosphere such that Earth was exposed mostly to interstellar GCRs and minimally to HEPs, {or to lower fluxes of HEPs}.

\section{Summary} \label{sec:summary}
We model the production of cosmogenic radioisotope $^{10}\text{Be}$ in Earth's atmosphere when the heliosphere crosses through an interstellar cold cloud. A cold-cloud encounter compresses the heliosphere and exposes Earth to increased radiation. We consider two cases where the heliosphere is compressed by clouds of different densities, such that Earth is exposed to pristine interstellar GCRs and HEPs{, which we compare to the cosmic ray exposure of a supernova occurring 50 pc from Earth}. We then consider a case where Earth is exposed only to interstellar GCRs. 

{If the heliosphere is compressed to 0.2 AU (such that the termination shock is compressed to 0.12 AU), the $^{10}\text{Be}$ production rate is increased by a factor of 9 compared to the background production rate from GCRs modulated in a non-compressed heliosphere at 1 AU. If the heliosphere is compressed to 0.7 AU (with a termination shock at 0.4 AU), the $^{10}\text{Be}$ production rate increases by a factor of 3. Interstellar GCRs alone increase the $^{10}\text{Be}$ production rate by less than a factor of 3.} {A supernova 50 pc away can increase $^{10}\text{Be}$ production by a factor of 4, with an elevated signal lasting $\sim75$ kyrs.}

% Depending on the amount of heliospheric compression, HEPs can increase the globally-averaged $^{10}\text{Be}$ production rate by a factor of {3-9} from a typical, non-compressed heliosphere. The ratio of HEP to GCR exposure can vary depending on the shape of the heliosphere and the relative direction between the Sun and the cloud, though exposure to HEPs for a minority of Earth's orbit substantially increases $^{10}\text{Be}$ production. 

Despite these increases in atmospheric $^{10}\text{Be}$ production, detectability of an interstellar cloud crossing is not guaranteed. Detectability in geologic records depends on the duration of the crossing, radiation exposure, and time resolutions of the samples. Encounters with interstellar clouds in our local neighborhood could be short-lived, since clouds within the Local Ribbon may be as thin as 200 AU \citep{meyer_remarkable_2012}. Some studies measuring cosmogenic isotopes {on Earth and the lunar surface} have not detected interstellar cloud crossings 2-3 or 6-7 Ma  \citep{poluianov_detectability_2025, koll_cosmogenic_2025, valet_geomagnetic_2025}. We find that methods with low temporal resolutions, such as sampling of Fe-Mn crusts {or lunar soil},  cannot detect sub-parsec-scale clouds or supernovae occurring mear Earth. {Interstellar cloud crossings which compress the heliosphere to sub-AU scales must last 0.5-1 Myr to be detected in Fe-Mn crusts.} Measurements of long-lived cosmogenic isotopes with high temporal resolutions, such as those in ocean sediments and ice cores, {depend on the duration of the crossing, the density of the cloud, and the resulting amount of heliospheric compression. Heliospheric compressions of 0.2 AU can be detected in ocean sediments and ice cores; less extreme heliospheric compressions or nearby supernovae cannot easily be distinguished in ocean sediments beyond terrestrial variability.} Our results suggest that the $^{10}\text{Be}$ peak seen by \cite{koll_cosmogenic_2025} {is too prolonged to be} attributed to a supernova; an interstellar cloud crossing may better fit this observation.

\section*{Availability Statement}
Code and model results are freely available at \cite{nica_data_2026}.

\section*{Conflict of Interest Statement}
The authors have no conflicts of interest to disclose.

\acknowledgments
{We thank Prof. Edward Brook for guidance regarding variability in ice core measurements.} This paper is based on work supported by NASA FINESST (grant number 80NSSC23K1638) and by NASA grant 18-DRIVE18\_2 as part of the SHIELD DRIVE program,``Our Heliospheric Shield" (80NSSC22M0164, \url{https://shielddrivecenter.com/}).
%%%%%%%%%%%%%%%%%%%%%%%%%%%%%%%%%%%%%%%%%%%%%%%
% REFERENCES and BIBLIOGRAPHYe
%
% \bibliography{<name of your .bib file>} don't specify the file extension
% don't specify bibliographystyle
%
%%%%%%%%%%%%%%%%%%%%%%%%%%%%%%%%%%%%%%%%%%%%%%%

% UNCOMMENT
\bibliography{agusample}

\begin{thebibliography}{}

\bibitem [\protect \citeauthoryear {%
{Adriani}%
\ \protect \BOthers {.}}{%
{Adriani}%
\ \protect \BOthers {.}}{%
{\protect \APACyear {2011}}%
}]{%
adriani_pamela_2011}
\APACinsertmetastar {%
adriani_pamela_2011}%
\begin{APACrefauthors}%
{Adriani}, O.%
, {Barbarino}, G\BPBI C.%
, {Bazilevskaya}, G\BPBI A.%
, {Bellotti}, R.%
, {Boezio}, M.%
, {Bogomolov}, E\BPBI A.%
\BDBL {}{Zverev}, V\BPBI G.%
\end{APACrefauthors}%
\unskip\
\newblock
\APACrefYearMonthDay{2011}{{\APACmonth{04}}}{}.
\newblock
{\BBOQ}\APACrefatitle {{PAMELA Measurements of Cosmic-Ray Proton and Helium Spectra}} {{PAMELA Measurements of Cosmic-Ray Proton and Helium Spectra}}.{\BBCQ}
\newblock
\APACjournalVolNumPages{Science}{332}{6025}{69}.
\newblock
\begin{APACrefDOI} \doi{10.1126/science.1199172} \end{APACrefDOI}
\PrintBackRefs{\CurrentBib}

\bibitem [\protect \citeauthoryear {%
Agostinelli%
\ \protect \BOthers {.}}{%
Agostinelli%
\ \protect \BOthers {.}}{%
{\protect \APACyear {2003}}%
}]{%
agostinelli_geant4simulation_2003}
\APACinsertmetastar {%
agostinelli_geant4simulation_2003}%
\begin{APACrefauthors}%
Agostinelli, S.%
, Allison, J.%
, Amako, K.%
, Apostolakis, J.%
, Araujo, H.%
, Arce, P.%
\BDBL {}Zschiesche, D.%
\end{APACrefauthors}%
\unskip\
\newblock
\APACrefYearMonthDay{2003}{{\APACmonth{07}}}{}.
\newblock
{\BBOQ}\APACrefatitle {Geant4—a simulation toolkit} {Geant4—a simulation toolkit}.{\BBCQ}
\newblock
\APACjournalVolNumPages{Nuclear Instruments and Methods in Physics Research Section A: Accelerators, Spectrometers, Detectors and Associated Equipment}{506}{3}{250--303}.
\newblock
\begin{APACrefURL} [{2025-08-25}]\url{https://www.sciencedirect.com/science/article/pii/S0168900203013688} \end{APACrefURL}
\newblock
\begin{APACrefDOI} \doi{10.1016/S0168-9002(03)01368-8} \end{APACrefDOI}
\PrintBackRefs{\CurrentBib}

\bibitem [\protect \citeauthoryear {%
Allison%
\ \protect \BOthers {.}}{%
Allison%
\ \protect \BOthers {.}}{%
{\protect \APACyear {2006}}%
}]{%
allison_geant4_2006}
\APACinsertmetastar {%
allison_geant4_2006}%
\begin{APACrefauthors}%
Allison, J.%
, Amako, K.%
, Apostolakis, J.%
, Araujo, H.%
, Arce~Dubois, P.%
, Asai, M.%
\BDBL {}Yoshida, H.%
\end{APACrefauthors}%
\unskip\
\newblock
\APACrefYearMonthDay{2006}{{\APACmonth{02}}}{}.
\newblock
{\BBOQ}\APACrefatitle {Geant4 developments and applications} {Geant4 developments and applications}.{\BBCQ}
\newblock
\APACjournalVolNumPages{IEEE Transactions on Nuclear Science}{53}{1}{270--278}.
\newblock
\begin{APACrefURL} [{2025-08-25}]\url{https://ieeexplore.ieee.org/document/1610988} \end{APACrefURL}
\newblock
\begin{APACrefDOI} \doi{10.1109/TNS.2006.869826} \end{APACrefDOI}
\PrintBackRefs{\CurrentBib}

\bibitem [\protect \citeauthoryear {%
{Bard}%
\ \protect \BOthers {.}}{%
{Bard}%
\ \protect \BOthers {.}}{%
{\protect \APACyear {2023}}%
}]{%
bard_radiocarbon_2023}
\APACinsertmetastar {%
bard_radiocarbon_2023}%
\begin{APACrefauthors}%
{Bard}, E.%
, {Miramont}, C.%
, {Capano}, M.%
, {Guibal}, F.%
, {Marschal}, C.%
, {Rostek}, F.%
\BDBL {}{Heaton}, T\BPBI J.%
\end{APACrefauthors}%
\unskip\
\newblock
\APACrefYearMonthDay{2023}{{\APACmonth{11}}}{}.
\newblock
{\BBOQ}\APACrefatitle {{A radiocarbon spike at 14 300 cal yr BP in subfossil trees provides the impulse response function of the global carbon cycle during the Late Glacial}} {{A radiocarbon spike at 14 300 cal yr BP in subfossil trees provides the impulse response function of the global carbon cycle during the Late Glacial}}.{\BBCQ}
\newblock
\APACjournalVolNumPages{Philosophical Transactions of the Royal Society of London Series A}{381}{2261}{20220206}.
\newblock
\begin{APACrefDOI} \doi{10.1098/rsta.2022.0206} \end{APACrefDOI}
\PrintBackRefs{\CurrentBib}

\bibitem [\protect \citeauthoryear {%
{Beer}%
\ \protect \BOthers {.}}{%
{Beer}%
\ \protect \BOthers {.}}{%
{\protect \APACyear {1990}}%
}]{%
beer_use_1990}
\APACinsertmetastar {%
beer_use_1990}%
\begin{APACrefauthors}%
{Beer}, J.%
, {Blinov}, A.%
, {Bonani}, G.%
, {Finkel}, R\BPBI C.%
, {Hofmann}, H\BPBI J.%
, {Lehmann}, B.%
\BDBL {}{W{\"o}tfli}, W.%
\end{APACrefauthors}%
\unskip\
\newblock
\APACrefYearMonthDay{1990}{{\APACmonth{09}}}{}.
\newblock
{\BBOQ}\APACrefatitle {{Use of $^{10}$Be in polar ice to trace the 11-year cycle of solar activity}} {{Use of $^{10}$Be in polar ice to trace the 11-year cycle of solar activity}}.{\BBCQ}
\newblock
\APACjournalVolNumPages{Nature}{347}{6289}{164-166}.
\newblock
\begin{APACrefDOI} \doi{10.1038/347164a0} \end{APACrefDOI}
\PrintBackRefs{\CurrentBib}

\bibitem [\protect \citeauthoryear {%
Biggin%
, McCormack%
\BCBL {}\ \BBA {} Roberts%
}{%
Biggin%
\ \protect \BOthers {.}}{%
{\protect \APACyear {2010}}%
}]{%
biggin_paleointensity_2010}
\APACinsertmetastar {%
biggin_paleointensity_2010}%
\begin{APACrefauthors}%
Biggin, A\BPBI J.%
, McCormack, A.%
\BCBL {}\ \BBA {} Roberts, A.%
\end{APACrefauthors}%
\unskip\
\newblock
\APACrefYearMonthDay{2010}{}{}.
\newblock
{\BBOQ}\APACrefatitle {Paleointensity {Database} {Updated} and {Upgraded}} {Paleointensity {Database} {Updated} and {Upgraded}}.{\BBCQ}
\newblock
\APACjournalVolNumPages{Eos, Transactions American Geophysical Union}{91}{2}{15--15}.
\newblock
\begin{APACrefURL} [{2025-10-22}]\url{https://onlinelibrary.wiley.com/doi/abs/10.1029/2010EO020003} \end{APACrefURL}
\newblock
\APACrefnote{\_eprint: https://agupubs.onlinelibrary.wiley.com/doi/pdf/10.1029/2010EO020003}
\newblock
\begin{APACrefDOI} \doi{10.1029/2010EO020003} \end{APACrefDOI}
\PrintBackRefs{\CurrentBib}

\bibitem [\protect \citeauthoryear {%
{Binns}%
\ \protect \BOthers {.}}{%
{Binns}%
\ \protect \BOthers {.}}{%
{\protect \APACyear {2016}}%
}]{%
binns_observation_2016}
\APACinsertmetastar {%
binns_observation_2016}%
\begin{APACrefauthors}%
{Binns}, W\BPBI R.%
, {Israel}, M\BPBI H.%
, {Christian}, E\BPBI R.%
, {Cummings}, A\BPBI C.%
, {de Nolfo}, G\BPBI A.%
, {Lave}, K\BPBI A.%
\BDBL {}{Wiedenbeck}, M\BPBI E.%
\end{APACrefauthors}%
\unskip\
\newblock
\APACrefYearMonthDay{2016}{{\APACmonth{05}}}{}.
\newblock
{\BBOQ}\APACrefatitle {{Observation of the $^{60}$Fe nucleosynthesis-clock isotope in galactic cosmic rays}} {{Observation of the $^{60}$Fe nucleosynthesis-clock isotope in galactic cosmic rays}}.{\BBCQ}
\newblock
\APACjournalVolNumPages{Science}{352}{6286}{677-680}.
\newblock
\begin{APACrefDOI} \doi{10.1126/science.aad6004} \end{APACrefDOI}
\PrintBackRefs{\CurrentBib}

\bibitem [\protect \citeauthoryear {%
Caballero-Lopez%
\ \BBA {} Moraal%
}{%
Caballero-Lopez%
\ \BBA {} Moraal%
}{%
{\protect \APACyear {2004}}%
}]{%
caballero-lopez_limitations_2004}
\APACinsertmetastar {%
caballero-lopez_limitations_2004}%
\begin{APACrefauthors}%
Caballero-Lopez, R\BPBI A.%
\BCBT {}\ \BBA {} Moraal, H.%
\end{APACrefauthors}%
\unskip\
\newblock
\APACrefYearMonthDay{2004}{}{}.
\newblock
{\BBOQ}\APACrefatitle {Limitations of the force field equation to describe cosmic ray modulation} {Limitations of the force field equation to describe cosmic ray modulation}.{\BBCQ}
\newblock
\APACjournalVolNumPages{Journal of Geophysical Research: Space Physics}{109}{A1}{}.
\newblock
\begin{APACrefURL} [{2024-07-25}]\url{https://onlinelibrary.wiley.com/doi/abs/10.1029/2003JA010098} \end{APACrefURL}
\newblock
\APACrefnote{\_eprint: https://onlinelibrary.wiley.com/doi/pdf/10.1029/2003JA010098}
\newblock
\begin{APACrefDOI} \doi{10.1029/2003JA010098} \end{APACrefDOI}
\PrintBackRefs{\CurrentBib}

\bibitem [\protect \citeauthoryear {%
{Cahlon}%
, {Zucker}%
, {Goodman}%
, {Lada}%
\BCBL {}\ \BBA {} {Alves}%
}{%
{Cahlon}%
\ \protect \BOthers {.}}{%
{\protect \APACyear {2024}}%
}]{%
cahlon_parsec_2024}
\APACinsertmetastar {%
cahlon_parsec_2024}%
\begin{APACrefauthors}%
{Cahlon}, S.%
, {Zucker}, C.%
, {Goodman}, A.%
, {Lada}, C.%
\BCBL {}\ \BBA {} {Alves}, J.%
\end{APACrefauthors}%
\unskip\
\newblock
\APACrefYearMonthDay{2024}{{\APACmonth{02}}}{}.
\newblock
{\BBOQ}\APACrefatitle {{A Parsec-scale Catalog of Molecular Clouds in the Solar Neighborhood Based on 3D Dust Mapping: Implications for the Mass-Size Relation}} {{A Parsec-scale Catalog of Molecular Clouds in the Solar Neighborhood Based on 3D Dust Mapping: Implications for the Mass-Size Relation}}.{\BBCQ}
\newblock
\APACjournalVolNumPages{The Astrophysical Journal}{961}{2}{153}.
\newblock
\begin{APACrefDOI} \doi{10.3847/1538-4357/ad0cf8} \end{APACrefDOI}
\PrintBackRefs{\CurrentBib}

\bibitem [\protect \citeauthoryear {%
Comedi%
, Elias%
\BCBL {}\ \BBA {} Zossi%
}{%
Comedi%
\ \protect \BOthers {.}}{%
{\protect \APACyear {2020}}%
}]{%
comedi_spatial_2020}
\APACinsertmetastar {%
comedi_spatial_2020}%
\begin{APACrefauthors}%
Comedi, E\BPBI S.%
, Elias, A\BPBI G.%
\BCBL {}\ \BBA {} Zossi, B\BPBI S.%
\end{APACrefauthors}%
\unskip\
\newblock
\APACrefYearMonthDay{2020}{{\APACmonth{12}}}{}.
\newblock
{\BBOQ}\APACrefatitle {Spatial features of geomagnetic cutoff rigidity secular variation using analytical approaches} {Spatial features of geomagnetic cutoff rigidity secular variation using analytical approaches}.{\BBCQ}
\newblock
\APACjournalVolNumPages{Journal of Atmospheric and Solar-Terrestrial Physics}{211}{}{105475}.
\newblock
\begin{APACrefURL} [{2024-08-22}]\url{https://www.sciencedirect.com/science/article/pii/S1364682620302789} \end{APACrefURL}
\newblock
\begin{APACrefDOI} \doi{10.1016/j.jastp.2020.105475} \end{APACrefDOI}
\PrintBackRefs{\CurrentBib}

\bibitem [\protect \citeauthoryear {%
Cummings%
\ \protect \BOthers {.}}{%
Cummings%
\ \protect \BOthers {.}}{%
{\protect \APACyear {2016}}%
}]{%
cummings_galactic_2016}
\APACinsertmetastar {%
cummings_galactic_2016}%
\begin{APACrefauthors}%
Cummings, A\BPBI C.%
, Stone, E\BPBI C.%
, Heikkila, B\BPBI C.%
, Lal, N.%
, Webber, W\BPBI R.%
, Jóhannesson, G.%
\BDBL {}Porter, T\BPBI A.%
\end{APACrefauthors}%
\unskip\
\newblock
\APACrefYearMonthDay{2016}{Nov}{}.
\newblock
{\BBOQ}\APACrefatitle {Galactic {Cosmic} {Rays} in the {Local} {Interstellar} {Medium}: {Voyager} 1 {Observations} and {Model} {Results}} {Galactic {Cosmic} {Rays} in the {Local} {Interstellar} {Medium}: {Voyager} 1 {Observations} and {Model} {Results}}.{\BBCQ}
\newblock
\APACjournalVolNumPages{The Astrophysical Journal}{831}{}{18}.
\newblock
\begin{APACrefURL} [{2024-05-14}]\url{https://ui.adsabs.harvard.edu/abs/2016ApJ...831...18C} \end{APACrefURL}
\newblock
\APACrefnote{Publisher: IOP ADS Bibcode: 2016ApJ...831...18C}
\newblock
\begin{APACrefDOI} \doi{10.3847/0004-637X/831/1/18} \end{APACrefDOI}
\PrintBackRefs{\CurrentBib}

\bibitem [\protect \citeauthoryear {%
{Decker}%
, {Krimigis}%
, {Roelof}%
\BCBL {}\ \BBA {} {Hill}%
}{%
{Decker}%
\ \protect \BOthers {.}}{%
{\protect \APACyear {2012}}%
}]{%
decker_no_2012}
\APACinsertmetastar {%
decker_no_2012}%
\begin{APACrefauthors}%
{Decker}, R\BPBI B.%
, {Krimigis}, S\BPBI M.%
, {Roelof}, E\BPBI C.%
\BCBL {}\ \BBA {} {Hill}, M\BPBI E.%
\end{APACrefauthors}%
\unskip\
\newblock
\APACrefYearMonthDay{2012}{{\APACmonth{09}}}{}.
\newblock
{\BBOQ}\APACrefatitle {{No meridional plasma flow in the heliosheath transition region}} {{No meridional plasma flow in the heliosheath transition region}}.{\BBCQ}
\newblock
\APACjournalVolNumPages{Nature}{489}{7414}{124-127}.
\newblock
\begin{APACrefDOI} \doi{10.1038/nature11441} \end{APACrefDOI}
\PrintBackRefs{\CurrentBib}

\bibitem [\protect \citeauthoryear {%
{Decker}%
\ \protect \BOthers {.}}{%
{Decker}%
\ \protect \BOthers {.}}{%
{\protect \APACyear {2008}}%
}]{%
decker_mediation_2008}
\APACinsertmetastar {%
decker_mediation_2008}%
\begin{APACrefauthors}%
{Decker}, R\BPBI B.%
, {Krimigis}, S\BPBI M.%
, {Roelof}, E\BPBI C.%
, {Hill}, M\BPBI E.%
, {Armstrong}, T\BPBI P.%
, {Gloeckler}, G.%
\BDBL {}{Lanzerotti}, L\BPBI J.%
\end{APACrefauthors}%
\unskip\
\newblock
\APACrefYearMonthDay{2008}{{\APACmonth{07}}}{}.
\newblock
{\BBOQ}\APACrefatitle {{Mediation of the solar wind termination shock by non-thermal ions}} {{Mediation of the solar wind termination shock by non-thermal ions}}.{\BBCQ}
\newblock
\APACjournalVolNumPages{Nature}{454}{7200}{67-70}.
\newblock
\begin{APACrefDOI} \doi{10.1038/nature07030} \end{APACrefDOI}
\PrintBackRefs{\CurrentBib}

\bibitem [\protect \citeauthoryear {%
{Delaygue}%
\ \BBA {} {Bard}%
}{%
{Delaygue}%
\ \BBA {} {Bard}%
}{%
{\protect \APACyear {2011}}%
}]{%
delaygue_antarctic_2011}
\APACinsertmetastar {%
delaygue_antarctic_2011}%
\begin{APACrefauthors}%
{Delaygue}, G.%
\BCBT {}\ \BBA {} {Bard}, E.%
\end{APACrefauthors}%
\unskip\
\newblock
\APACrefYearMonthDay{2011}{{\APACmonth{06}}}{}.
\newblock
{\BBOQ}\APACrefatitle {{An Antarctic view of Beryllium-10 and solar activity for the past millennium}} {{An Antarctic view of Beryllium-10 and solar activity for the past millennium}}.{\BBCQ}
\newblock
\APACjournalVolNumPages{Climate Dynamics}{36}{11-12}{2201-2218}.
\newblock
\begin{APACrefDOI} \doi{10.1007/s00382-010-0795-1} \end{APACrefDOI}
\PrintBackRefs{\CurrentBib}

\bibitem [\protect \citeauthoryear {%
{Ellison}%
\ \BBA {} {Ramaty}%
}{%
{Ellison}%
\ \BBA {} {Ramaty}%
}{%
{\protect \APACyear {1985}}%
}]{%
ellison_shock_1985}
\APACinsertmetastar {%
ellison_shock_1985}%
\begin{APACrefauthors}%
{Ellison}, D\BPBI C.%
\BCBT {}\ \BBA {} {Ramaty}, R.%
\end{APACrefauthors}%
\unskip\
\newblock
\APACrefYearMonthDay{1985}{{\APACmonth{11}}}{}.
\newblock
{\BBOQ}\APACrefatitle {{Shock acceleration of electrons and ions in solar flares}} {{Shock acceleration of electrons and ions in solar flares}}.{\BBCQ}
\newblock
\APACjournalVolNumPages{The Astrophysical Journal}{298}{}{400-408}.
\newblock
\begin{APACrefDOI} \doi{10.1086/163623} \end{APACrefDOI}
\PrintBackRefs{\CurrentBib}

\bibitem [\protect \citeauthoryear {%
{Fimiani}%
\ \protect \BOthers {.}}{%
{Fimiani}%
\ \protect \BOthers {.}}{%
{\protect \APACyear {2016}}%
}]{%
fimiani_interstellar_2016}
\APACinsertmetastar {%
fimiani_interstellar_2016}%
\begin{APACrefauthors}%
{Fimiani}, L.%
, {Cook}, D\BPBI L.%
, {Faestermann}, T.%
, {G{\'o}mez-Guzm{\'a}n}, J\BPBI M.%
, {Hain}, K.%
, {Herzog}, G.%
\BDBL {}{Rugel}, G.%
\end{APACrefauthors}%
\unskip\
\newblock
\APACrefYearMonthDay{2016}{{\APACmonth{04}}}{}.
\newblock
{\BBOQ}\APACrefatitle {{Interstellar <mml:mmultiscripts>Fe 60 </mml:mmultiscripts> on the Surface of the Moon}} {{Interstellar <mml:mmultiscripts>Fe 60 </mml:mmultiscripts> on the Surface of the Moon}}.{\BBCQ}
\newblock
\APACjournalVolNumPages{Physical Review Letters}{116}{15}{151104}.
\newblock
\begin{APACrefDOI} \doi{10.1103/PhysRevLett.116.151104} \end{APACrefDOI}
\PrintBackRefs{\CurrentBib}

\bibitem [\protect \citeauthoryear {%
{Fitoussi}%
\ \protect \BOthers {.}}{%
{Fitoussi}%
\ \protect \BOthers {.}}{%
{\protect \APACyear {2008}}%
}]{%
fitoussi_search_2008}
\APACinsertmetastar {%
fitoussi_search_2008}%
\begin{APACrefauthors}%
{Fitoussi}, C.%
, {Raisbeck}, G\BPBI M.%
, {Knie}, K.%
, {Korschinek}, G.%
, {Faestermann}, T.%
, {Goriely}, S.%
\BDBL {}{Wallner}, A.%
\end{APACrefauthors}%
\unskip\
\newblock
\APACrefYearMonthDay{2008}{{\APACmonth{09}}}{}.
\newblock
{\BBOQ}\APACrefatitle {{Search for Supernova-Produced Fe60 in a Marine Sediment}} {{Search for Supernova-Produced Fe60 in a Marine Sediment}}.{\BBCQ}
\newblock
\APACjournalVolNumPages{Physical Review Letters}{101}{12}{121101}.
\newblock
\begin{APACrefDOI} \doi{10.1103/PhysRevLett.101.121101} \end{APACrefDOI}
\PrintBackRefs{\CurrentBib}

\bibitem [\protect \citeauthoryear {%
Frank%
\ \protect \BOthers {.}}{%
Frank%
\ \protect \BOthers {.}}{%
{\protect \APACyear {1997}}%
}]{%
frank_200_1997}
\APACinsertmetastar {%
frank_200_1997}%
\begin{APACrefauthors}%
Frank, M.%
, Schwarz, B.%
, Baumann, S.%
, Kubik, P\BPBI W.%
, Suter, M.%
\BCBL {}\ \BBA {} Mangini, A.%
\end{APACrefauthors}%
\unskip\
\newblock
\APACrefYearMonthDay{1997}{{\APACmonth{06}}}{}.
\newblock
{\BBOQ}\APACrefatitle {A 200 kyr record of cosmogenic radionuclide production rate and geomagnetic field intensity from {10Be} in globally stacked deep-sea sediments1} {A 200 kyr record of cosmogenic radionuclide production rate and geomagnetic field intensity from {10Be} in globally stacked deep-sea sediments1}.{\BBCQ}
\newblock
\APACjournalVolNumPages{Earth and Planetary Science Letters}{149}{1}{121--129}.
\newblock
\begin{APACrefURL} [{2025-09-15}]\url{https://www.sciencedirect.com/science/article/pii/S0012821X97000708} \end{APACrefURL}
\newblock
\begin{APACrefDOI} \doi{10.1016/S0012-821X(97)00070-8} \end{APACrefDOI}
\PrintBackRefs{\CurrentBib}

\bibitem [\protect \citeauthoryear {%
{Giacalone}%
\ \protect \BOthers {.}}{%
{Giacalone}%
\ \protect \BOthers {.}}{%
{\protect \APACyear {2022}}%
}]{%
giacalone_anomalous_2022}
\APACinsertmetastar {%
giacalone_anomalous_2022}%
\begin{APACrefauthors}%
{Giacalone}, J.%
, {Fahr}, H.%
, {Fichtner}, H.%
, {Florinski}, V.%
, {Heber}, B.%
, {Hill}, M\BPBI E.%
\BDBL {}{Rankin}, J\BPBI S.%
\end{APACrefauthors}%
\unskip\
\newblock
\APACrefYearMonthDay{2022}{{\APACmonth{06}}}{}.
\newblock
{\BBOQ}\APACrefatitle {{Anomalous Cosmic Rays and Heliospheric Energetic Particles}} {{Anomalous Cosmic Rays and Heliospheric Energetic Particles}}.{\BBCQ}
\newblock
\APACjournalVolNumPages{Space Science Reviews}{218}{4}{22}.
\newblock
\begin{APACrefDOI} \doi{10.1007/s11214-022-00890-7} \end{APACrefDOI}
\PrintBackRefs{\CurrentBib}

\bibitem [\protect \citeauthoryear {%
Golubenko%
\ \protect \BOthers {.}}{%
Golubenko%
\ \protect \BOthers {.}}{%
{\protect \APACyear {2024}}%
}]{%
golubenko_full_2024}
\APACinsertmetastar {%
golubenko_full_2024}%
\begin{APACrefauthors}%
Golubenko, K.%
, Rozanov, E.%
, Kovaltsov, G.%
, Baroni, M.%
, Sukhodolov, T.%
\BCBL {}\ \BBA {} Usoskin, I.%
\end{APACrefauthors}%
\unskip\
\newblock
\APACrefYearMonthDay{2024}{}{}.
\newblock
{\BBOQ}\APACrefatitle {Full {Modeling} and {Practical} {Parameterization} of {Cosmogenic} {10Be} {Transport} for {Cosmic}-{Ray} {Studies}: {SOCOL}-{AERv2}-{BE} {Model}} {Full {Modeling} and {Practical} {Parameterization} of {Cosmogenic} {10Be} {Transport} for {Cosmic}-{Ray} {Studies}: {SOCOL}-{AERv2}-{BE} {Model}}.{\BBCQ}
\newblock
\APACjournalVolNumPages{Journal of Geophysical Research: Space Physics}{129}{7}{e2024JA032504}.
\newblock
\begin{APACrefURL} [{2025-02-19}]\url{https://onlinelibrary.wiley.com/doi/abs/10.1029/2024JA032504} \end{APACrefURL}
\newblock
\APACrefnote{\_eprint: https://onlinelibrary.wiley.com/doi/pdf/10.1029/2024JA032504}
\newblock
\begin{APACrefDOI} \doi{10.1029/2024JA032504} \end{APACrefDOI}
\PrintBackRefs{\CurrentBib}

\bibitem [\protect \citeauthoryear {%
Golubenko%
\ \protect \BOthers {.}}{%
Golubenko%
\ \protect \BOthers {.}}{%
{\protect \APACyear {2021}}%
}]{%
golubenko_application_2021}
\APACinsertmetastar {%
golubenko_application_2021}%
\begin{APACrefauthors}%
Golubenko, K.%
, Rozanov, E.%
, Kovaltsov, G.%
, Leppänen, A\BHBI P.%
, Sukhodolov, T.%
\BCBL {}\ \BBA {} Usoskin, I.%
\end{APACrefauthors}%
\unskip\
\newblock
\APACrefYearMonthDay{2021}{{\APACmonth{12}}}{}.
\newblock
{\BBOQ}\APACrefatitle {Application of {CCM} {SOCOL}-{AERv2}-{BE} to cosmogenic beryllium isotopes: description and validation for polar regions} {Application of {CCM} {SOCOL}-{AERv2}-{BE} to cosmogenic beryllium isotopes: description and validation for polar regions}.{\BBCQ}
\newblock
\APACjournalVolNumPages{Geoscientific Model Development}{14}{12}{7605--7620}.
\newblock
\begin{APACrefURL} [{2025-06-06}]\url{https://gmd.copernicus.org/articles/14/7605/2021/} \end{APACrefURL}
\newblock
\APACrefnote{Publisher: Copernicus GmbH}
\newblock
\begin{APACrefDOI} \doi{10.5194/gmd-14-7605-2021} \end{APACrefDOI}
\PrintBackRefs{\CurrentBib}

\bibitem [\protect \citeauthoryear {%
Heikkilä%
, Beer%
\BCBL {}\ \BBA {} Feichter%
}{%
Heikkilä%
\ \protect \BOthers {.}}{%
{\protect \APACyear {2009}}%
}]{%
heikkila_meridional_2009}
\APACinsertmetastar {%
heikkila_meridional_2009}%
\begin{APACrefauthors}%
Heikkilä, U.%
, Beer, J.%
\BCBL {}\ \BBA {} Feichter, J.%
\end{APACrefauthors}%
\unskip\
\newblock
\APACrefYearMonthDay{2009}{{\APACmonth{01}}}{}.
\newblock
{\BBOQ}\APACrefatitle {Meridional transport and deposition of atmospheric $^{\textrm{10}}${Be}} {Meridional transport and deposition of atmospheric $^{\textrm{10}}${Be}}.{\BBCQ}
\newblock
\APACjournalVolNumPages{Atmospheric Chemistry and Physics}{9}{2}{515--527}.
\newblock
\begin{APACrefURL} [{2025-01-27}]\url{https://acp.copernicus.org/articles/9/515/2009/acp-9-515-2009.html} \end{APACrefURL}
\newblock
\APACrefnote{Publisher: Copernicus GmbH}
\newblock
\begin{APACrefDOI} \doi{10.5194/acp-9-515-2009} \end{APACrefDOI}
\PrintBackRefs{\CurrentBib}

\bibitem [\protect \citeauthoryear {%
Hein%
\ \protect \BOthers {.}}{%
Hein%
\ \protect \BOthers {.}}{%
{\protect \APACyear {2000}}%
}]{%
hein_cobalt-rich_2000}
\APACinsertmetastar {%
hein_cobalt-rich_2000}%
\begin{APACrefauthors}%
Hein, J\BPBI R.%
, Koschinsk, A.%
, Bau, M.%
, Manheim, F\BPBI T.%
, Kang, J\BHBI K.%
\BCBL {}\ \BBA {} Roberts, L.%
\end{APACrefauthors}%
\unskip\
\newblock
\APACrefYearMonthDay{2000}{}{}.
\newblock
{\BBOQ}\APACrefatitle {Cobalt-{Rich} {Ferromanganese} {Crusts} in the {Pacific}} {Cobalt-{Rich} {Ferromanganese} {Crusts} in the {Pacific}}.{\BBCQ}
\newblock
\BIn{} \APACrefbtitle {Handbook of {Marine} {Mineral} {Deposits}.} {Handbook of {Marine} {Mineral} {Deposits}.}
\newblock
\APACaddressPublisher{}{Routledge}.
\newblock
\APACrefnote{Num Pages: 41}
\PrintBackRefs{\CurrentBib}

\bibitem [\protect \citeauthoryear {%
Hein%
\ \protect \BOthers {.}}{%
Hein%
\ \protect \BOthers {.}}{%
{\protect \APACyear {1997}}%
}]{%
hein_iron_1997}
\APACinsertmetastar {%
hein_iron_1997}%
\begin{APACrefauthors}%
Hein, J\BPBI R.%
, Koschinsky, A.%
, Halbach, P.%
, Manheim, F\BPBI T.%
, Bau, M.%
, Kang, J\BHBI K.%
\BCBL {}\ \BBA {} Lubick, N.%
\end{APACrefauthors}%
\unskip\
\newblock
\APACrefYearMonthDay{1997}{{\APACmonth{01}}}{}.
\newblock
{\BBOQ}\APACrefatitle {Iron and manganese oxide mineralization in the {Pacific}} {Iron and manganese oxide mineralization in the {Pacific}}.{\BBCQ}
\newblock
\APACjournalVolNumPages{Geological Society, London, Special Publications}{119}{1}{123--138}.
\newblock
\begin{APACrefURL} [{2026-02-05}]\url{https://www.lyellcollection.org/doi/abs/10.1144/GSL.SP.1997.119.01.09} \end{APACrefURL}
\newblock
\begin{APACrefDOI} \doi{10.1144/GSL.SP.1997.119.01.09} \end{APACrefDOI}
\PrintBackRefs{\CurrentBib}

\bibitem [\protect \citeauthoryear {%
{Knie}%
\ \protect \BOthers {.}}{%
{Knie}%
\ \protect \BOthers {.}}{%
{\protect \APACyear {2004}}%
}]{%
knie_60Fe_2004}
\APACinsertmetastar {%
knie_60Fe_2004}%
\begin{APACrefauthors}%
{Knie}, K.%
, {Korschinek}, G.%
, {Faestermann}, T.%
, {Dorfi}, E\BPBI A.%
, {Rugel}, G.%
\BCBL {}\ \BBA {} {Wallner}, A.%
\end{APACrefauthors}%
\unskip\
\newblock
\APACrefYearMonthDay{2004}{{\APACmonth{10}}}{}.
\newblock
{\BBOQ}\APACrefatitle {{$^{60}$Fe Anomaly in a Deep-Sea Manganese Crust and Implications for a Nearby Supernova Source}} {{$^{60}$Fe Anomaly in a Deep-Sea Manganese Crust and Implications for a Nearby Supernova Source}}.{\BBCQ}
\newblock
\APACjournalVolNumPages{Physical Review Letters}{93}{17}{171103}.
\newblock
\begin{APACrefDOI} \doi{10.1103/PhysRevLett.93.171103} \end{APACrefDOI}
\PrintBackRefs{\CurrentBib}

\bibitem [\protect \citeauthoryear {%
{Knie}%
\ \protect \BOthers {.}}{%
{Knie}%
\ \protect \BOthers {.}}{%
{\protect \APACyear {1999}}%
}]{%
knie_indication_1999}
\APACinsertmetastar {%
knie_indication_1999}%
\begin{APACrefauthors}%
{Knie}, K.%
, {Korschinek}, G.%
, {Faestermann}, T.%
, {Wallner}, C.%
, {Scholten}, J.%
\BCBL {}\ \BBA {} {Hillebrandt}, W.%
\end{APACrefauthors}%
\unskip\
\newblock
\APACrefYearMonthDay{1999}{{\APACmonth{07}}}{}.
\newblock
{\BBOQ}\APACrefatitle {{Indication for Supernova Produced $^{60}$Fe Activity on Earth}} {{Indication for Supernova Produced $^{60}$Fe Activity on Earth}}.{\BBCQ}
\newblock
\APACjournalVolNumPages{Physical Review Letters}{83}{1}{18-21}.
\newblock
\begin{APACrefDOI} \doi{10.1103/PhysRevLett.83.18} \end{APACrefDOI}
\PrintBackRefs{\CurrentBib}

\bibitem [\protect \citeauthoryear {%
{Koll}%
, {Feige}%
, {Lachner}%
\BCBL {}\ \BBA {} {Wallner}%
}{%
{Koll}%
\ \protect \BOthers {.}}{%
{\protect \APACyear {2026}}%
}]{%
koll_no_2026}
\APACinsertmetastar {%
koll_no_2026}%
\begin{APACrefauthors}%
{Koll}, D.%
, {Feige}, J.%
, {Lachner}, J.%
\BCBL {}\ \BBA {} {Wallner}, A.%
\end{APACrefauthors}%
\unskip\
\newblock
\APACrefYearMonthDay{2026}{{\APACmonth{05}}}{}.
\newblock
{\BBOQ}\APACrefatitle {{No indication of a strong increase in galactic cosmic ray intensity 2-3 Myr ago from cosmogenic nuclides}} {{No indication of a strong increase in galactic cosmic ray intensity 2-3 Myr ago from cosmogenic nuclides}}.{\BBCQ}
\newblock
\APACjournalVolNumPages{Nature Astronomy}{10}{}{630-632}.
\newblock
\begin{APACrefDOI} \doi{10.1038/s41550-026-02834-5} \end{APACrefDOI}
\PrintBackRefs{\CurrentBib}

\bibitem [\protect \citeauthoryear {%
Koll%
\ \protect \BOthers {.}}{%
Koll%
\ \protect \BOthers {.}}{%
{\protect \APACyear {2025}}%
}]{%
koll_cosmogenic_2025}
\APACinsertmetastar {%
koll_cosmogenic_2025}%
\begin{APACrefauthors}%
Koll, D.%
, Lachner, J.%
, Beutner, S.%
, Fichter, S.%
, Merchel, S.%
, Rugel, G.%
\BDBL {}Wallner, A.%
\end{APACrefauthors}%
\unskip\
\newblock
\APACrefYearMonthDay{2025}{{\APACmonth{02}}}{}.
\newblock
{\BBOQ}\APACrefatitle {A cosmogenic {10Be} anomaly during the late {Miocene} as independent time marker for marine archives} {A cosmogenic {10Be} anomaly during the late {Miocene} as independent time marker for marine archives}.{\BBCQ}
\newblock
\APACjournalVolNumPages{Nature Communications}{16}{1}{866}.
\newblock
\begin{APACrefURL} [{2025-02-17}]\url{https://www.nature.com/articles/s41467-024-55662-4} \end{APACrefURL}
\newblock
\APACrefnote{Publisher: Nature Publishing Group}
\newblock
\begin{APACrefDOI} \doi{10.1038/s41467-024-55662-4} \end{APACrefDOI}
\PrintBackRefs{\CurrentBib}

\bibitem [\protect \citeauthoryear {%
{Krimigis}%
\ \protect \BOthers {.}}{%
{Krimigis}%
\ \protect \BOthers {.}}{%
{\protect \APACyear {2019}}%
}]{%
krimigis_energetic_2019}
\APACinsertmetastar {%
krimigis_energetic_2019}%
\begin{APACrefauthors}%
{Krimigis}, S\BPBI M.%
, {Decker}, R\BPBI B.%
, {Roelof}, E\BPBI C.%
, {Hill}, M\BPBI E.%
, {Bostrom}, C\BPBI O.%
, {Dialynas}, K.%
\BDBL {}{Lanzerotti}, L\BPBI J.%
\end{APACrefauthors}%
\unskip\
\newblock
\APACrefYearMonthDay{2019}{{\APACmonth{11}}}{}.
\newblock
{\BBOQ}\APACrefatitle {{Energetic charged particle measurements from Voyager 2 at the heliopause and beyond}} {{Energetic charged particle measurements from Voyager 2 at the heliopause and beyond}}.{\BBCQ}
\newblock
\APACjournalVolNumPages{Nature Astronomy}{3}{}{997-1006}.
\newblock
\begin{APACrefDOI} \doi{10.1038/s41550-019-0927-4} \end{APACrefDOI}
\PrintBackRefs{\CurrentBib}

\bibitem [\protect \citeauthoryear {%
Lal%
\ \BBA {} Peters%
}{%
Lal%
\ \BBA {} Peters%
}{%
{\protect \APACyear {1967}}%
}]{%
lal_cosmic_1967}
\APACinsertmetastar {%
lal_cosmic_1967}%
\begin{APACrefauthors}%
Lal, D.%
\BCBT {}\ \BBA {} Peters, B.%
\end{APACrefauthors}%
\unskip\
\newblock
\APACrefYearMonthDay{1967}{}{}.
\newblock
{\BBOQ}\APACrefatitle {Cosmic {Ray} {Produced} {Radioactivity} on the {Earth}} {Cosmic {Ray} {Produced} {Radioactivity} on the {Earth}}.{\BBCQ}
\newblock
\BIn{} K.~Sitte\ (\BED), \APACrefbtitle {Kosmische {Strahlung} {II} / {Cosmic} {Rays} {II}} {Kosmische {Strahlung} {II} / {Cosmic} {Rays} {II}}\ (\BPGS\ 551--612).
\newblock
\APACaddressPublisher{Berlin, Heidelberg}{Springer}.
\newblock
\begin{APACrefURL} [{2025-10-22}]\url{https://doi.org/10.1007/978-3-642-46079-1_7} \end{APACrefURL}
\newblock
\begin{APACrefDOI} \doi{10.1007/978-3-642-46079-1_7} \end{APACrefDOI}
\PrintBackRefs{\CurrentBib}

\bibitem [\protect \citeauthoryear {%
{Ludwig}%
\ \protect \BOthers {.}}{%
{Ludwig}%
\ \protect \BOthers {.}}{%
{\protect \APACyear {2016}}%
}]{%
ludwig_time_2016}
\APACinsertmetastar {%
ludwig_time_2016}%
\begin{APACrefauthors}%
{Ludwig}, P.%
, {Bishop}, S.%
, {Egli}, R.%
, {Chernenko}, V.%
, {Deneva}, B.%
, {Faestermann}, T.%
\BDBL {}{Rugel}, G.%
\end{APACrefauthors}%
\unskip\
\newblock
\APACrefYearMonthDay{2016}{{\APACmonth{08}}}{}.
\newblock
{\BBOQ}\APACrefatitle {{Time-resolved 2-million-year-old supernova activity discovered in Earth's microfossil record}} {{Time-resolved 2-million-year-old supernova activity discovered in Earth's microfossil record}}.{\BBCQ}
\newblock
\APACjournalVolNumPages{Proceedings of the National Academy of Science}{113}{33}{9232-9237}.
\newblock
\begin{APACrefDOI} \doi{10.1073/pnas.1601040113} \end{APACrefDOI}
\PrintBackRefs{\CurrentBib}

\bibitem [\protect \citeauthoryear {%
{McDonald}%
}{%
{McDonald}%
}{%
{\protect \APACyear {1998}}%
}]{%
mcdonald_cosmic_1998}
\APACinsertmetastar {%
mcdonald_cosmic_1998}%
\begin{APACrefauthors}%
{McDonald}, F\BPBI B.%
\end{APACrefauthors}%
\unskip\
\newblock
\APACrefYearMonthDay{1998}{{\APACmonth{01}}}{}.
\newblock
{\BBOQ}\APACrefatitle {{Cosmic-Ray Modulation in the Heliosphere A Phenomenological Study}} {{Cosmic-Ray Modulation in the Heliosphere A Phenomenological Study}}.{\BBCQ}
\newblock
\APACjournalVolNumPages{Space Science Reviews}{83}{}{33-50}.
\PrintBackRefs{\CurrentBib}

\bibitem [\protect \citeauthoryear {%
{Mekhaldi}%
\ \protect \BOthers {.}}{%
{Mekhaldi}%
\ \protect \BOthers {.}}{%
{\protect \APACyear {2015}}%
}]{%
mekhaldi_multiradionuclide_2015}
\APACinsertmetastar {%
mekhaldi_multiradionuclide_2015}%
\begin{APACrefauthors}%
{Mekhaldi}, F.%
, {Muscheler}, R.%
, {Adolphi}, F.%
, {Aldahan}, A.%
, {Beer}, J.%
, {McConnell}, J\BPBI R.%
\BDBL {}{Woodruff}, T\BPBI E.%
\end{APACrefauthors}%
\unskip\
\newblock
\APACrefYearMonthDay{2015}{{\APACmonth{10}}}{}.
\newblock
{\BBOQ}\APACrefatitle {{Multiradionuclide evidence for the solar origin of the cosmic-ray events of AD 774/5 and 993/4}} {{Multiradionuclide evidence for the solar origin of the cosmic-ray events of AD 774/5 and 993/4}}.{\BBCQ}
\newblock
\APACjournalVolNumPages{Nature Communications}{6}{}{8611}.
\newblock
\begin{APACrefDOI} \doi{10.1038/ncomms9611} \end{APACrefDOI}
\PrintBackRefs{\CurrentBib}

\bibitem [\protect \citeauthoryear {%
{Mewaldt}%
\ \protect \BOthers {.}}{%
{Mewaldt}%
\ \protect \BOthers {.}}{%
{\protect \APACyear {2005}}%
}]{%
mewaldt_proton_2005}
\APACinsertmetastar {%
mewaldt_proton_2005}%
\begin{APACrefauthors}%
{Mewaldt}, R\BPBI A.%
, {Cohen}, C\BPBI M\BPBI S.%
, {Labrador}, A\BPBI W.%
, {Leske}, R\BPBI A.%
, {Mason}, G\BPBI M.%
, {Desai}, M\BPBI I.%
\BDBL {}{Haggerty}, D\BPBI K.%
\end{APACrefauthors}%
\unskip\
\newblock
\APACrefYearMonthDay{2005}{{\APACmonth{09}}}{}.
\newblock
{\BBOQ}\APACrefatitle {{Proton, helium, and electron spectra during the large solar particle events of October-November 2003}} {{Proton, helium, and electron spectra during the large solar particle events of October-November 2003}}.{\BBCQ}
\newblock
\APACjournalVolNumPages{Journal of Geophysical Research (Space Physics)}{110}{A9}{A09S18}.
\newblock
\begin{APACrefDOI} \doi{10.1029/2005JA011038} \end{APACrefDOI}
\PrintBackRefs{\CurrentBib}

\bibitem [\protect \citeauthoryear {%
Meyer%
, Lauroesch%
, Peek%
\BCBL {}\ \BBA {} Heiles%
}{%
Meyer%
\ \protect \BOthers {.}}{%
{\protect \APACyear {2012}}%
}]{%
meyer_remarkable_2012}
\APACinsertmetastar {%
meyer_remarkable_2012}%
\begin{APACrefauthors}%
Meyer, D\BPBI M.%
, Lauroesch, J\BPBI T.%
, Peek, J\BPBI E\BPBI G.%
\BCBL {}\ \BBA {} Heiles, C.%
\end{APACrefauthors}%
\unskip\
\newblock
\APACrefYearMonthDay{2012}{{\APACmonth{06}}}{}.
\newblock
{\BBOQ}\APACrefatitle {{The} {Remarkable} {High} {Pressure} {of} {the} {Local} {Leo} {Cold} {Cloud}} {{The} {Remarkable} {High} {Pressure} {of} {the} {Local} {Leo} {Cold} {Cloud}}.{\BBCQ}
\newblock
\APACjournalVolNumPages{The Astrophysical Journal}{752}{2}{119}.
\newblock
\begin{APACrefURL} [{2025-03-12}]\url{https://dx.doi.org/10.1088/0004-637X/752/2/119} \end{APACrefURL}
\newblock
\APACrefnote{Publisher: The American Astronomical Society}
\newblock
\begin{APACrefDOI} \doi{10.1088/0004-637X/752/2/119} \end{APACrefDOI}
\PrintBackRefs{\CurrentBib}

\bibitem [\protect \citeauthoryear {%
{Middleton}%
\ \protect \BOthers {.}}{%
{Middleton}%
\ \protect \BOthers {.}}{%
{\protect \APACyear {2026}}%
}]{%
middleton_oceanographic_2026}
\APACinsertmetastar {%
middleton_oceanographic_2026}%
\begin{APACrefauthors}%
{Middleton}, J.%
, {Pavia}, F.%
, {Anderson}, R.%
, {Schwartz}, R.%
, {Fleisher}, M.%
, {Lao}, Y.%
\BDBL {}{Winckler}, G.%
\end{APACrefauthors}%
\unskip\
\newblock
\APACrefYearMonthDay{2026}{{\APACmonth{07}}}{}.
\newblock
{\BBOQ}\APACrefatitle {{Oceanographic and climatic controls of meteoric $^{10}$Be fluxes to seafloor sediments: A global synthesis}} {{Oceanographic and climatic controls of meteoric $^{10}$Be fluxes to seafloor sediments: A global synthesis}}.{\BBCQ}
\newblock
\APACjournalVolNumPages{Quaternary Science Reviews}{383}{}{109922}.
\newblock
\begin{APACrefDOI} \doi{10.1016/j.quascirev.2026.109922} \end{APACrefDOI}
\PrintBackRefs{\CurrentBib}

\bibitem [\protect \citeauthoryear {%
Miyake%
\ \protect \BOthers {.}}{%
Miyake%
\ \protect \BOthers {.}}{%
{\protect \APACyear {2015}}%
}]{%
miyake_cosmic_2015}
\APACinsertmetastar {%
miyake_cosmic_2015}%
\begin{APACrefauthors}%
Miyake, F.%
, Suzuki, A.%
, Masuda, K.%
, Horiuchi, K.%
, Motoyama, H.%
, Matsuzaki, H.%
\BDBL {}Nakai, Y.%
\end{APACrefauthors}%
\unskip\
\newblock
\APACrefYearMonthDay{2015}{}{}.
\newblock
{\BBOQ}\APACrefatitle {Cosmic ray event of A.D. 774–775 shown in quasi-annual 10Be data from the Antarctic Dome Fuji ice core} {Cosmic ray event of a.d. 774–775 shown in quasi-annual 10be data from the antarctic dome fuji ice core}.{\BBCQ}
\newblock
\APACjournalVolNumPages{Geophysical Research Letters}{42}{1}{84-89}.
\newblock
\begin{APACrefURL} \url{https://agupubs-onlinelibrary-wiley-com.ezproxy.bu.edu/doi/abs/10.1002/2014GL062218} \end{APACrefURL}
\newblock
\begin{APACrefDOI} \doi{https://doi-org.ezproxy.bu.edu/10.1002/2014GL062218} \end{APACrefDOI}
\PrintBackRefs{\CurrentBib}

\bibitem [\protect \citeauthoryear {%
Miyake%
, Usoskin%
\BCBL {}\ \BBA {} Poluianov%
}{%
Miyake%
\ \protect \BOthers {.}}{%
{\protect \APACyear {2019}}%
}]{%
miyake_extreme_2019}
\APACinsertmetastar {%
miyake_extreme_2019}%
\begin{APACrefauthors}%
Miyake, F.%
, Usoskin, I.%
\BCBL {}\ \BBA {} Poluianov, S.%
\end{APACrefauthors}%
\ (\BEDS).
\unskip\
\newblock
\APACrefYear{2019}.
\newblock
\APACrefbtitle {Extreme {Solar} {Particle} {Storms}} {Extreme {Solar} {Particle} {Storms}}.
\newblock
\APACaddressPublisher{}{IOP Publishing}.
\newblock
\begin{APACrefURL} \url{https://doi.org/10.1088/2514-3433/ab404a} \end{APACrefURL}
\newblock
\begin{APACrefDOI} \doi{10.1088/2514-3433/ab404a} \end{APACrefDOI}
\PrintBackRefs{\CurrentBib}

\bibitem [\protect \citeauthoryear {%
Morlino%
\ \BBA {} Gabici%
}{%
Morlino%
\ \BBA {} Gabici%
}{%
{\protect \APACyear {2015}}%
}]{%
morlino_cosmic_2015}
\APACinsertmetastar {%
morlino_cosmic_2015}%
\begin{APACrefauthors}%
Morlino, G.%
\BCBT {}\ \BBA {} Gabici, S.%
\end{APACrefauthors}%
\unskip\
\newblock
\APACrefYearMonthDay{2015}{{\APACmonth{07}}}{}.
\newblock
{\BBOQ}\APACrefatitle {Cosmic ray penetration in diffuse clouds} {Cosmic ray penetration in diffuse clouds}.{\BBCQ}
\newblock
\APACjournalVolNumPages{Monthly Notices of the Royal Astronomical Society: Letters}{451}{1}{L100--L104}.
\newblock
\begin{APACrefURL} [{2025-07-02}]\url{https://doi.org/10.1093/mnrasl/slv074} \end{APACrefURL}
\newblock
\begin{APACrefDOI} \doi{10.1093/mnrasl/slv074} \end{APACrefDOI}
\PrintBackRefs{\CurrentBib}

\bibitem [\protect \citeauthoryear {%
Nica%
, Opher%
, Miller%
, Middleton%
\BCBL {}\ \BBA {} Giacalone%
}{%
Nica%
\ \protect \BOthers {.}}{%
{\protect \APACyear {2026}}%
}]{%
nica_data_2026}
\APACinsertmetastar {%
nica_data_2026}%
\begin{APACrefauthors}%
Nica, A.%
, Opher, M.%
, Miller, J.%
, Middleton, J.%
\BCBL {}\ \BBA {} Giacalone, J.%
\end{APACrefauthors}%
\unskip\
\newblock
\APACrefYearMonthDay{2026}{{\APACmonth{03}}}{}.
\newblock
\APACrefbtitle {Data {Supporting} {Nica} et al. "{Cosmogenic} {10Be} as a {Tracer} for {Recent} {Heliospheric} {Encounters} with {Interstellar} {Cold} {Clouds}".} {Data {Supporting} {Nica} et al. "{Cosmogenic} {10Be} as a {Tracer} for {Recent} {Heliospheric} {Encounters} with {Interstellar} {Cold} {Clouds}".}
\newblock
\APACaddressPublisher{}{Zenodo}.
\newblock
\begin{APACrefURL} \url{https://doi.org/10.5281/zenodo.21630922} \end{APACrefURL}
\newblock
\begin{APACrefDOI} \doi{10.5281/zenodo.21630922} \end{APACrefDOI}
\PrintBackRefs{\CurrentBib}

\bibitem [\protect \citeauthoryear {%
{O'Hare}%
\ \protect \BOthers {.}}{%
{O'Hare}%
\ \protect \BOthers {.}}{%
{\protect \APACyear {2019}}%
}]{%
ohare_multiradionuclide_2019}
\APACinsertmetastar {%
ohare_multiradionuclide_2019}%
\begin{APACrefauthors}%
{O'Hare}, P.%
, {Mekhaldi}, F.%
, {Adolphi}, F.%
, {Raisbeck}, G.%
, {Aldahan}, A.%
, {Anderberg}, E.%
\BDBL {}et al.%
\end{APACrefauthors}%
\unskip\
\newblock
\APACrefYearMonthDay{2019}{{\APACmonth{03}}}{}.
\newblock
{\BBOQ}\APACrefatitle {{Multiradionuclide evidence for an extreme solar proton event around 2,610 B.P. ({\ensuremath{\sim}}660 BC)}} {{Multiradionuclide evidence for an extreme solar proton event around 2,610 B.P. ({\ensuremath{\sim}}660 BC)}}.{\BBCQ}
\newblock
\APACjournalVolNumPages{Proceedings of the National Academy of Science}{116}{13}{5961-5966}.
\newblock
\begin{APACrefDOI} \doi{10.1073/pnas.1815725116} \end{APACrefDOI}
\PrintBackRefs{\CurrentBib}

\bibitem [\protect \citeauthoryear {%
Opher%
\ \protect \BOthers {.}}{%
Opher%
\ \protect \BOthers {.}}{%
{\protect \APACyear {2026}}%
}]{%
opher_increased_2026}
\APACinsertmetastar {%
opher_increased_2026}%
\begin{APACrefauthors}%
Opher, M.%
, Giacalone, J.%
, Loeb, A.%
, Economo, E\BPBI P.%
, Cummings, A.%
, Middleton, J.%
\BDBL {}Hatzaki, M.%
\end{APACrefauthors}%
\unskip\
\newblock
\APACrefYearMonthDay{2026}{{\APACmonth{02}}}{}.
\newblock
{\BBOQ}\APACrefatitle {Increased and varied radiation during the {Sun}’s encounters with cold clouds in the last 10 million years} {Increased and varied radiation during the {Sun}’s encounters with cold clouds in the last 10 million years}.{\BBCQ}
\newblock
\APACjournalVolNumPages{Scientific Reports}{}{}{}.
\newblock
\begin{APACrefURL} [{2026-02-19}]\url{https://www.nature.com/articles/s41598-026-36926-z} \end{APACrefURL}
\newblock
\begin{APACrefDOI} \doi{10.1038/s41598-026-36926-z} \end{APACrefDOI}
\PrintBackRefs{\CurrentBib}

\bibitem [\protect \citeauthoryear {%
{Opher}%
\ \protect \BOthers {.}}{%
{Opher}%
\ \protect \BOthers {.}}{%
{\protect \APACyear {2026}}%
}]{%
opher_reply_2025}
\APACinsertmetastar {%
opher_reply_2025}%
\begin{APACrefauthors}%
{Opher}, M.%
, {Loeb}, A.%
, {Peek}, J\BPBI E.%
, {Miller}, J.%
, {Middleton}, J.%
\BCBL {}\ \BBA {} {Nica}, A.%
\end{APACrefauthors}%
\unskip\
\newblock
\APACrefYearMonthDay{2026}{{\APACmonth{05}}}{}.
\newblock
{\BBOQ}\APACrefatitle {{Reply to: No indication of a strong increase in galactic cosmic ray intensity 2-3 Myr ago from cosmogenic nuclides}} {{Reply to: No indication of a strong increase in galactic cosmic ray intensity 2-3 Myr ago from cosmogenic nuclides}}.{\BBCQ}
\newblock
\APACjournalVolNumPages{Nature Astronomy}{10}{}{633-635}.
\newblock
\begin{APACrefDOI} \doi{10.1038/s41550-026-02835-4} \end{APACrefDOI}
\PrintBackRefs{\CurrentBib}

\bibitem [\protect \citeauthoryear {%
Opher%
, Loeb%
\BCBL {}\ \BBA {} Peek%
}{%
Opher%
, Loeb%
\BCBL {}\ \BBA {} Peek%
}{%
{\protect \APACyear {2024}}%
}]{%
opher_possible_2024}
\APACinsertmetastar {%
opher_possible_2024}%
\begin{APACrefauthors}%
Opher, M.%
, Loeb, A.%
\BCBL {}\ \BBA {} Peek, J\BPBI E\BPBI G.%
\end{APACrefauthors}%
\unskip\
\newblock
\APACrefYearMonthDay{2024}{{\APACmonth{06}}}{}.
\newblock
{\BBOQ}\APACrefatitle {A possible direct exposure of the {Earth} to the cold dense interstellar medium 2–3 {Myr} ago} {A possible direct exposure of the {Earth} to the cold dense interstellar medium 2–3 {Myr} ago}.{\BBCQ}
\newblock
\APACjournalVolNumPages{Nature Astronomy}{}{}{1--8}.
\newblock
\begin{APACrefURL} [{2024-06-10}]\url{https://www.nature.com/articles/s41550-024-02279-8} \end{APACrefURL}
\newblock
\APACrefnote{Publisher: Nature Publishing Group}
\newblock
\begin{APACrefDOI} \doi{10.1038/s41550-024-02279-8} \end{APACrefDOI}
\PrintBackRefs{\CurrentBib}

\bibitem [\protect \citeauthoryear {%
Opher%
, Loeb%
, Zucker%
\BCBL {}\ \protect \BOthers {.}}{%
Opher%
, Loeb%
, Zucker%
\BCBL {}\ \protect \BOthers {.}}{%
{\protect \APACyear {2024}}%
}]{%
opher_passage_2024}
\APACinsertmetastar {%
opher_passage_2024}%
\begin{APACrefauthors}%
Opher, M.%
, Loeb, A.%
, Zucker, C.%
, Goodman, A.%
, Konietzka, R.%
, Worden, A\BPBI Z.%
\BDBL {}Foley, M\BPBI M.%
\end{APACrefauthors}%
\unskip\
\newblock
\APACrefYearMonthDay{2024}{{\APACmonth{09}}}{}.
\newblock
{\BBOQ}\APACrefatitle {The {Passage} of the {Solar} {System} through the {Edge} of the {Local} {Bubble}} {The {Passage} of the {Solar} {System} through the {Edge} of the {Local} {Bubble}}.{\BBCQ}
\newblock
\APACjournalVolNumPages{The Astrophysical Journal}{972}{2}{201}.
\newblock
\begin{APACrefURL} [{2025-02-11}]\url{https://dx.doi.org/10.3847/1538-4357/ad596e} \end{APACrefURL}
\newblock
\APACrefnote{Publisher: The American Astronomical Society}
\newblock
\begin{APACrefDOI} \doi{10.3847/1538-4357/ad596e} \end{APACrefDOI}
\PrintBackRefs{\CurrentBib}

\bibitem [\protect \citeauthoryear {%
{Paleari}%
\ \protect \BOthers {.}}{%
{Paleari}%
\ \protect \BOthers {.}}{%
{\protect \APACyear {2022}}%
}]{%
paleari_cosmogenic_2022}
\APACinsertmetastar {%
paleari_cosmogenic_2022}%
\begin{APACrefauthors}%
{Paleari}, C\BPBI I.%
, {Mekhaldi}, F.%
, {Adolphi}, F.%
, {Christl}, M.%
, {Vockenhuber}, C.%
, {Gautschi}, P.%
\BDBL {}et al.%
\end{APACrefauthors}%
\unskip\
\newblock
\APACrefYearMonthDay{2022}{{\APACmonth{01}}}{}.
\newblock
{\BBOQ}\APACrefatitle {{Cosmogenic radionuclides reveal an extreme solar particle storm near a solar minimum 9125 years BP}} {{Cosmogenic radionuclides reveal an extreme solar particle storm near a solar minimum 9125 years BP}}.{\BBCQ}
\newblock
\APACjournalVolNumPages{Nature Communications}{13}{}{214}.
\newblock
\begin{APACrefDOI} \doi{10.1038/s41467-021-27891-4} \end{APACrefDOI}
\PrintBackRefs{\CurrentBib}

\bibitem [\protect \citeauthoryear {%
Poluianov%
\ \BBA {} Engelbrecht%
}{%
Poluianov%
\ \BBA {} Engelbrecht%
}{%
{\protect \APACyear {2025}}%
}]{%
poluianov_detectability_2025}
\APACinsertmetastar {%
poluianov_detectability_2025}%
\begin{APACrefauthors}%
Poluianov, S.%
\BCBT {}\ \BBA {} Engelbrecht, N\BPBI E.%
\end{APACrefauthors}%
\unskip\
\newblock
\APACrefYearMonthDay{2025}{{\APACmonth{02}}}{}.
\newblock
{\BBOQ}\APACrefatitle {Detectability of the passage of the heliosphere through an interstellar cloud with cosmogenic nuclides in lunar soil} {Detectability of the passage of the heliosphere through an interstellar cloud with cosmogenic nuclides in lunar soil}.{\BBCQ}
\newblock
\APACjournalVolNumPages{Astronomy \& Astrophysics}{694}{}{A62}.
\newblock
\begin{APACrefURL} [{2025-02-07}]\url{https://www.aanda.org/10.1051/0004-6361/202452337} \end{APACrefURL}
\newblock
\begin{APACrefDOI} \doi{10.1051/0004-6361/202452337} \end{APACrefDOI}
\PrintBackRefs{\CurrentBib}

\bibitem [\protect \citeauthoryear {%
Poluianov%
, Kovaltsov%
, Mishev%
\BCBL {}\ \BBA {} Usoskin%
}{%
Poluianov%
\ \protect \BOthers {.}}{%
{\protect \APACyear {2016}}%
}]{%
poluianov_production_2016}
\APACinsertmetastar {%
poluianov_production_2016}%
\begin{APACrefauthors}%
Poluianov, S.%
, Kovaltsov, G\BPBI A.%
, Mishev, A\BPBI L.%
\BCBL {}\ \BBA {} Usoskin, I\BPBI G.%
\end{APACrefauthors}%
\unskip\
\newblock
\APACrefYearMonthDay{2016}{}{}.
\newblock
{\BBOQ}\APACrefatitle {Production of cosmogenic isotopes {7Be}, {10Be}, {14C}, {22Na}, and {36Cl} in the atmosphere: {Altitudinal} profiles of yield functions} {Production of cosmogenic isotopes {7Be}, {10Be}, {14C}, {22Na}, and {36Cl} in the atmosphere: {Altitudinal} profiles of yield functions}.{\BBCQ}
\newblock
\APACjournalVolNumPages{Journal of Geophysical Research: Atmospheres}{121}{13}{8125--8136}.
\newblock
\begin{APACrefURL} [{2025-01-07}]\url{https://onlinelibrary.wiley.com/doi/abs/10.1002/2016JD025034} \end{APACrefURL}
\newblock
\APACrefnote{\_eprint: https://onlinelibrary.wiley.com/doi/pdf/10.1002/2016JD025034}
\newblock
\begin{APACrefDOI} \doi{10.1002/2016JD025034} \end{APACrefDOI}
\PrintBackRefs{\CurrentBib}

\bibitem [\protect \citeauthoryear {%
Poluianov%
, Kovaltsov%
\BCBL {}\ \BBA {} Usoskin%
}{%
Poluianov%
\ \protect \BOthers {.}}{%
{\protect \APACyear {2018}}%
}]{%
poluianov_solar_2018}
\APACinsertmetastar {%
poluianov_solar_2018}%
\begin{APACrefauthors}%
Poluianov, S.%
, Kovaltsov, G\BPBI A.%
\BCBL {}\ \BBA {} Usoskin, I\BPBI G.%
\end{APACrefauthors}%
\unskip\
\newblock
\APACrefYearMonthDay{2018}{{\APACmonth{10}}}{}.
\newblock
{\BBOQ}\APACrefatitle {Solar energetic particles and galactic cosmic rays over millions of years as inferred from data on cosmogenic {26Al} in lunar samples} {Solar energetic particles and galactic cosmic rays over millions of years as inferred from data on cosmogenic {26Al} in lunar samples}.{\BBCQ}
\newblock
\APACjournalVolNumPages{Astronomy \& Astrophysics}{618}{}{A96}.
\newblock
\begin{APACrefURL} [{2025-08-25}]\url{https://www.aanda.org/articles/aa/abs/2018/10/aa33561-18/aa33561-18.html} \end{APACrefURL}
\newblock
\APACrefnote{Publisher: EDP Sciences}
\newblock
\begin{APACrefDOI} \doi{10.1051/0004-6361/201833561} \end{APACrefDOI}
\PrintBackRefs{\CurrentBib}

\bibitem [\protect \citeauthoryear {%
{Raisbeck}%
, {Yiou}%
, {Cattani}%
\BCBL {}\ \BBA {} {Jouzel}%
}{%
{Raisbeck}%
\ \protect \BOthers {.}}{%
{\protect \APACyear {2006}}%
}]{%
raisbeck_evidence_2006}
\APACinsertmetastar {%
raisbeck_evidence_2006}%
\begin{APACrefauthors}%
{Raisbeck}, G\BPBI M.%
, {Yiou}, F.%
, {Cattani}, O.%
\BCBL {}\ \BBA {} {Jouzel}, J.%
\end{APACrefauthors}%
\unskip\
\newblock
\APACrefYearMonthDay{2006}{{\APACmonth{11}}}{}.
\newblock
{\BBOQ}\APACrefatitle {{$^{10}$Be evidence for the Matuyama-Brunhes geomagnetic reversal in the EPICA Dome C ice core}} {{$^{10}$Be evidence for the Matuyama-Brunhes geomagnetic reversal in the EPICA Dome C ice core}}.{\BBCQ}
\newblock
\APACjournalVolNumPages{Nature}{444}{7115}{82-84}.
\newblock
\begin{APACrefDOI} \doi{10.1038/nature05266} \end{APACrefDOI}
\PrintBackRefs{\CurrentBib}

\bibitem [\protect \citeauthoryear {%
{Raisbeck}%
, {Yiou}%
, {Jouzel}%
\BCBL {}\ \BBA {} {Stocker}%
}{%
{Raisbeck}%
\ \protect \BOthers {.}}{%
{\protect \APACyear {2007}}%
}]{%
raisbeck_direct_2007}
\APACinsertmetastar {%
raisbeck_direct_2007}%
\begin{APACrefauthors}%
{Raisbeck}, G\BPBI M.%
, {Yiou}, F.%
, {Jouzel}, J.%
\BCBL {}\ \BBA {} {Stocker}, T\BPBI F.%
\end{APACrefauthors}%
\unskip\
\newblock
\APACrefYearMonthDay{2007}{{\APACmonth{09}}}{}.
\newblock
{\BBOQ}\APACrefatitle {{Direct north-south synchronization of abrupt climate change record in ice cores using Beryllium 10}} {{Direct north-south synchronization of abrupt climate change record in ice cores using Beryllium 10}}.{\BBCQ}
\newblock
\APACjournalVolNumPages{Climate of the Past}{3}{3}{541-547}.
\newblock
\begin{APACrefDOI} \doi{10.5194/cp-3-541-200710.5194/cpd-3-755-2007} \end{APACrefDOI}
\PrintBackRefs{\CurrentBib}

\bibitem [\protect \citeauthoryear {%
{Richardson}%
, {Kasper}%
, {Wang}%
, {Belcher}%
\BCBL {}\ \BBA {} {Lazarus}%
}{%
{Richardson}%
\ \protect \BOthers {.}}{%
{\protect \APACyear {2008}}%
}]{%
richardson_cool_2008}
\APACinsertmetastar {%
richardson_cool_2008}%
\begin{APACrefauthors}%
{Richardson}, J\BPBI D.%
, {Kasper}, J\BPBI C.%
, {Wang}, C.%
, {Belcher}, J\BPBI W.%
\BCBL {}\ \BBA {} {Lazarus}, A\BPBI J.%
\end{APACrefauthors}%
\unskip\
\newblock
\APACrefYearMonthDay{2008}{{\APACmonth{07}}}{}.
\newblock
{\BBOQ}\APACrefatitle {{Cool heliosheath plasma and deceleration of the upstream solar wind at the termination shock}} {{Cool heliosheath plasma and deceleration of the upstream solar wind at the termination shock}}.{\BBCQ}
\newblock
\APACjournalVolNumPages{Nature}{454}{7200}{63-66}.
\newblock
\begin{APACrefDOI} \doi{10.1038/nature07024} \end{APACrefDOI}
\PrintBackRefs{\CurrentBib}

\bibitem [\protect \citeauthoryear {%
{Savranskaia}%
\ \protect \BOthers {.}}{%
{Savranskaia}%
\ \protect \BOthers {.}}{%
{\protect \APACyear {2024}}%
}]{%
savranskaia_removing_2024}
\APACinsertmetastar {%
savranskaia_removing_2024}%
\begin{APACrefauthors}%
{Savranskaia}, T.%
, {Egli}, R.%
, {Simon}, Q.%
, {Valet}, J\BHBI P.%
, {Bassinot}, F.%
\BCBL {}\ \BBA {} {Thouveny}, N.%
\end{APACrefauthors}%
\unskip\
\newblock
\APACrefYearMonthDay{2024}{{\APACmonth{12}}}{}.
\newblock
{\BBOQ}\APACrefatitle {{Removing Climatic Overprints in Sedimentary Cosmogenic Beryllium Records: Potentials and Limits}} {{Removing Climatic Overprints in Sedimentary Cosmogenic Beryllium Records: Potentials and Limits}}.{\BBCQ}
\newblock
\APACjournalVolNumPages{Geochemistry, Geophysics, Geosystems}{25}{12}{2024GC011761}.
\newblock
\begin{APACrefDOI} \doi{10.1029/2024GC011761} \end{APACrefDOI}
\PrintBackRefs{\CurrentBib}

\bibitem [\protect \citeauthoryear {%
{Savranskaia}%
\ \protect \BOthers {.}}{%
{Savranskaia}%
\ \protect \BOthers {.}}{%
{\protect \APACyear {2021}}%
}]{%
savranskaia_disentangling_2021}
\APACinsertmetastar {%
savranskaia_disentangling_2021}%
\begin{APACrefauthors}%
{Savranskaia}, T.%
, {Egli}, R.%
, {Valet}, J\BHBI P.%
, {Bassinot}, F.%
, {Meynadier}, L.%
, {Bourl{\`e}s}, D\BPBI L.%
\BDBL {}{Thouveny}, N.%
\end{APACrefauthors}%
\unskip\
\newblock
\APACrefYearMonthDay{2021}{{\APACmonth{04}}}{}.
\newblock
{\BBOQ}\APACrefatitle {{Disentangling magnetic and environmental signatures of sedimentary $^{10}$Be/$^{9}$Be records}} {{Disentangling magnetic and environmental signatures of sedimentary $^{10}$Be/$^{9}$Be records}}.{\BBCQ}
\newblock
\APACjournalVolNumPages{Quaternary Science Reviews}{257}{}{106809}.
\newblock
\begin{APACrefDOI} \doi{10.1016/j.quascirev.2021.106809} \end{APACrefDOI}
\PrintBackRefs{\CurrentBib}

\bibitem [\protect \citeauthoryear {%
Shackleton%
\ \protect \BOthers {.}}{%
Shackleton%
\ \protect \BOthers {.}}{%
{\protect \APACyear {2025}}%
}]{%
shackleton_miocene_2025}
\APACinsertmetastar {%
shackleton_miocene_2025}%
\begin{APACrefauthors}%
Shackleton, S.%
, Hishamunda, V.%
, Davidge, L.%
, Brook, E.%
, Peterson, J\BPBI M.%
, Carter, A.%
\BDBL {}Higgins, J\BPBI A.%
\end{APACrefauthors}%
\unskip\
\newblock
\APACrefYearMonthDay{2025}{}{}.
\newblock
{\BBOQ}\APACrefatitle {Miocene and Pliocene ice and air from the Allan Hills blue ice area, East Antarctica} {Miocene and pliocene ice and air from the allan hills blue ice area, east antarctica}.{\BBCQ}
\newblock
\APACjournalVolNumPages{Proceedings of the National Academy of Sciences}{122}{44}{e2502681122}.
\newblock
\begin{APACrefURL} \url{https://www.pnas.org/doi/abs/10.1073/pnas.2502681122} \end{APACrefURL}
\newblock
\begin{APACrefDOI} \doi{10.1073/pnas.2502681122} \end{APACrefDOI}
\PrintBackRefs{\CurrentBib}

\bibitem [\protect \citeauthoryear {%
Simon%
\ \protect \BOthers {.}}{%
Simon%
\ \protect \BOthers {.}}{%
{\protect \APACyear {2018}}%
}]{%
simon_increased_2018}
\APACinsertmetastar {%
simon_increased_2018}%
\begin{APACrefauthors}%
Simon, Q.%
, Thouveny, N.%
, Bourlès, D\BPBI L.%
, Bassinot, F.%
, Savranskaia, T.%
\BCBL {}\ \BBA {} Valet, J\BHBI P.%
\end{APACrefauthors}%
\unskip\
\newblock
\APACrefYearMonthDay{2018}{{\APACmonth{05}}}{}.
\newblock
{\BBOQ}\APACrefatitle {Increased production of cosmogenic {10Be} recorded in oceanic sediment sequences: {Information} on the age, duration, and amplitude of the geomagnetic dipole moment minimum over the {Matuyama}–{Brunhes} transition} {Increased production of cosmogenic {10Be} recorded in oceanic sediment sequences: {Information} on the age, duration, and amplitude of the geomagnetic dipole moment minimum over the {Matuyama}–{Brunhes} transition}.{\BBCQ}
\newblock
\APACjournalVolNumPages{Earth and Planetary Science Letters}{489}{}{191--202}.
\newblock
\begin{APACrefURL} [{2025-09-15}]\url{https://www.sciencedirect.com/science/article/pii/S0012821X18300980} \end{APACrefURL}
\newblock
\begin{APACrefDOI} \doi{10.1016/j.epsl.2018.02.036} \end{APACrefDOI}
\PrintBackRefs{\CurrentBib}

\bibitem [\protect \citeauthoryear {%
Stone%
, Cummings%
, Heikkila%
\BCBL {}\ \BBA {} Lal%
}{%
Stone%
\ \protect \BOthers {.}}{%
{\protect \APACyear {2019}}%
}]{%
stone_cosmic_2019}
\APACinsertmetastar {%
stone_cosmic_2019}%
\begin{APACrefauthors}%
Stone, E\BPBI C.%
, Cummings, A\BPBI C.%
, Heikkila, B\BPBI C.%
\BCBL {}\ \BBA {} Lal, N.%
\end{APACrefauthors}%
\unskip\
\newblock
\APACrefYearMonthDay{2019}{Nov}{}.
\newblock
{\BBOQ}\APACrefatitle {Cosmic ray measurements from {Voyager} 2 as it crossed into interstellar space} {Cosmic ray measurements from {Voyager} 2 as it crossed into interstellar space}.{\BBCQ}
\newblock
\APACjournalVolNumPages{Nature Astronomy}{3}{11}{1013--1018}.
\newblock
\begin{APACrefURL} [{2024-07-10}]\url{https://www.nature.com/articles/s41550-019-0928-3} \end{APACrefURL}
\newblock
\APACrefnote{Publisher: Nature Publishing Group}
\newblock
\begin{APACrefDOI} \doi{10.1038/s41550-019-0928-3} \end{APACrefDOI}
\PrintBackRefs{\CurrentBib}

\bibitem [\protect \citeauthoryear {%
{Stone}%
\ \protect \BOthers {.}}{%
{Stone}%
\ \protect \BOthers {.}}{%
{\protect \APACyear {2005}}%
}]{%
stone_voyager_2005}
\APACinsertmetastar {%
stone_voyager_2005}%
\begin{APACrefauthors}%
{Stone}, E\BPBI C.%
, {Cummings}, A\BPBI C.%
, {McDonald}, F\BPBI B.%
, {Heikkila}, B\BPBI C.%
, {Lal}, N.%
\BCBL {}\ \BBA {} {Webber}, W\BPBI R.%
\end{APACrefauthors}%
\unskip\
\newblock
\APACrefYearMonthDay{2005}{{\APACmonth{09}}}{}.
\newblock
{\BBOQ}\APACrefatitle {{Voyager 1 Explores the Termination Shock Region and the Heliosheath Beyond}} {{Voyager 1 Explores the Termination Shock Region and the Heliosheath Beyond}}.{\BBCQ}
\newblock
\APACjournalVolNumPages{Science}{309}{5743}{2017-2020}.
\newblock
\begin{APACrefDOI} \doi{10.1126/science.1117684} \end{APACrefDOI}
\PrintBackRefs{\CurrentBib}

\bibitem [\protect \citeauthoryear {%
{Stone}%
\ \protect \BOthers {.}}{%
{Stone}%
\ \protect \BOthers {.}}{%
{\protect \APACyear {2008}}%
}]{%
stone_asymmetric_2008}
\APACinsertmetastar {%
stone_asymmetric_2008}%
\begin{APACrefauthors}%
{Stone}, E\BPBI C.%
, {Cummings}, A\BPBI C.%
, {McDonald}, F\BPBI B.%
, {Heikkila}, B\BPBI C.%
, {Lal}, N.%
\BCBL {}\ \BBA {} {Webber}, W\BPBI R.%
\end{APACrefauthors}%
\unskip\
\newblock
\APACrefYearMonthDay{2008}{{\APACmonth{07}}}{}.
\newblock
{\BBOQ}\APACrefatitle {{An asymmetric solar wind termination shock}} {{An asymmetric solar wind termination shock}}.{\BBCQ}
\newblock
\APACjournalVolNumPages{Nature}{454}{7200}{71-74}.
\newblock
\begin{APACrefDOI} \doi{10.1038/nature07022} \end{APACrefDOI}
\PrintBackRefs{\CurrentBib}

\bibitem [\protect \citeauthoryear {%
Stone%
\ \protect \BOthers {.}}{%
Stone%
\ \protect \BOthers {.}}{%
{\protect \APACyear {2013}}%
}]{%
stone_voyager_2013}
\APACinsertmetastar {%
stone_voyager_2013}%
\begin{APACrefauthors}%
Stone, E\BPBI C.%
, Cummings, A\BPBI C.%
, McDonald, F\BPBI B.%
, Heikkila, B\BPBI C.%
, Lal, N.%
\BCBL {}\ \BBA {} Webber, W\BPBI R.%
\end{APACrefauthors}%
\unskip\
\newblock
\APACrefYearMonthDay{2013}{{\APACmonth{07}}}{}.
\newblock
{\BBOQ}\APACrefatitle {Voyager 1 {Observes} {Low}-{Energy} {Galactic} {Cosmic} {Rays} in a {Region} {Depleted} of {Heliospheric} {Ions}} {Voyager 1 {Observes} {Low}-{Energy} {Galactic} {Cosmic} {Rays} in a {Region} {Depleted} of {Heliospheric} {Ions}}.{\BBCQ}
\newblock
\APACjournalVolNumPages{Science}{341}{6142}{150--153}.
\newblock
\begin{APACrefURL} [{2024-07-10}]\url{https://www.science.org/doi/10.1126/science.1236408} \end{APACrefURL}
\newblock
\APACrefnote{Publisher: American Association for the Advancement of Science}
\newblock
\begin{APACrefDOI} \doi{10.1126/science.1236408} \end{APACrefDOI}
\PrintBackRefs{\CurrentBib}

\bibitem [\protect \citeauthoryear {%
{Thomas}%
\ \BBA {} {Yelland}%
}{%
{Thomas}%
\ \BBA {} {Yelland}%
}{%
{\protect \APACyear {2023}}%
}]{%
thomas_terrestrial_2023}
\APACinsertmetastar {%
thomas_terrestrial_2023}%
\begin{APACrefauthors}%
{Thomas}, B\BPBI C.%
\BCBT {}\ \BBA {} {Yelland}, A\BPBI M.%
\end{APACrefauthors}%
\unskip\
\newblock
\APACrefYearMonthDay{2023}{{\APACmonth{06}}}{}.
\newblock
{\BBOQ}\APACrefatitle {{Terrestrial Effects of Nearby Supernovae: Updated Modeling}} {{Terrestrial Effects of Nearby Supernovae: Updated Modeling}}.{\BBCQ}
\newblock
\APACjournalVolNumPages{The Astrophysical Journal}{950}{1}{41}.
\newblock
\begin{APACrefDOI} \doi{10.3847/1538-4357/accf8a} \end{APACrefDOI}
\PrintBackRefs{\CurrentBib}

\bibitem [\protect \citeauthoryear {%
{Valet}%
\ \protect \BOthers {.}}{%
{Valet}%
\ \protect \BOthers {.}}{%
{\protect \APACyear {2025}}%
}]{%
valet_geomagnetic_2025}
\APACinsertmetastar {%
valet_geomagnetic_2025}%
\begin{APACrefauthors}%
{Valet}, J\BHBI P.%
, {Savranskaia}, T.%
, {Simon}, Q.%
, {Egli}, R.%
, {Bassinot}, F.%
\BCBL {}\ \BBA {} {Thouveny}, N.%
\end{APACrefauthors}%
\unskip\
\newblock
\APACrefYearMonthDay{2025}{{\APACmonth{07}}}{}.
\newblock
{\BBOQ}\APACrefatitle {{Geomagnetic moment variation over the last 4.4 Million years through high resolution paleointensity and $^{10}$Be production records along ocean sediment sequences}} {{Geomagnetic moment variation over the last 4.4 Million years through high resolution paleointensity and $^{10}$Be production records along ocean sediment sequences}}.{\BBCQ}
\newblock
\APACjournalVolNumPages{Quaternary Science Reviews}{359}{}{109367}.
\newblock
\begin{APACrefDOI} \doi{10.1016/j.quascirev.2025.109367} \end{APACrefDOI}
\PrintBackRefs{\CurrentBib}

\bibitem [\protect \citeauthoryear {%
Vannier%
\ \protect \BOthers {.}}{%
Vannier%
\ \protect \BOthers {.}}{%
{\protect \APACyear {2025}}%
}]{%
vannier_mapping_2025}
\APACinsertmetastar {%
vannier_mapping_2025}%
\begin{APACrefauthors}%
Vannier, H.%
, Redfield, S.%
, Wood, B\BPBI E.%
, Müller, H\BHBI R.%
, Linsky, J\BPBI L.%
\BCBL {}\ \BBA {} Frisch, P\BPBI C.%
\end{APACrefauthors}%
\unskip\
\newblock
\APACrefYearMonthDay{2025}{{\APACmonth{03}}}{}.
\newblock
{\BBOQ}\APACrefatitle {Mapping {Our} {Path} through the {Local} {Interstellar} {Medium}: {High}-resolution {Ultraviolet} {Absorption} {Spectroscopy} of {Sight} {Lines} along the {Sun}’s {Historical} {Trajectory}} {Mapping {Our} {Path} through the {Local} {Interstellar} {Medium}: {High}-resolution {Ultraviolet} {Absorption} {Spectroscopy} of {Sight} {Lines} along the {Sun}’s {Historical} {Trajectory}}.{\BBCQ}
\newblock
\APACjournalVolNumPages{The Astrophysical Journal}{981}{2}{102}.
\newblock
\begin{APACrefURL} [{2025-09-11}]\url{https://dx.doi.org/10.3847/1538-4357/adb033} \end{APACrefURL}
\newblock
\APACrefnote{Publisher: The American Astronomical Society}
\newblock
\begin{APACrefDOI} \doi{10.3847/1538-4357/adb033} \end{APACrefDOI}
\PrintBackRefs{\CurrentBib}

\bibitem [\protect \citeauthoryear {%
Vos%
\ \BBA {} Potgieter%
}{%
Vos%
\ \BBA {} Potgieter%
}{%
{\protect \APACyear {2015}}%
}]{%
vos_new_2015}
\APACinsertmetastar {%
vos_new_2015}%
\begin{APACrefauthors}%
Vos, E\BPBI E.%
\BCBT {}\ \BBA {} Potgieter, M\BPBI S.%
\end{APACrefauthors}%
\unskip\
\newblock
\APACrefYearMonthDay{2015}{{\APACmonth{12}}}{}.
\newblock
{\BBOQ}\APACrefatitle {New {Modeling} of {Galactic} {Proton} {Modulation} {During} the {Minimum} of {Solar} {Cycle} 23/24} {New {Modeling} of {Galactic} {Proton} {Modulation} {During} the {Minimum} of {Solar} {Cycle} 23/24}.{\BBCQ}
\newblock
\APACjournalVolNumPages{The Astrophysical Journal}{815}{2}{119}.
\newblock
\begin{APACrefURL} [{2024-07-01}]\url{https://dx.doi.org/10.1088/0004-637X/815/2/119} \end{APACrefURL}
\newblock
\APACrefnote{Publisher: The American Astronomical Society}
\newblock
\begin{APACrefDOI} \doi{10.1088/0004-637X/815/2/119} \end{APACrefDOI}
\PrintBackRefs{\CurrentBib}

\bibitem [\protect \citeauthoryear {%
Wallner%
\ \protect \BOthers {.}}{%
Wallner%
\ \protect \BOthers {.}}{%
{\protect \APACyear {2016}}%
}]{%
wallner_recent_2016}
\APACinsertmetastar {%
wallner_recent_2016}%
\begin{APACrefauthors}%
Wallner, A.%
, Feige, J.%
, Kinoshita, N.%
, Paul, M.%
, Fifield, L\BPBI K.%
, Golser, R.%
\BDBL {}Winkler, S\BPBI R.%
\end{APACrefauthors}%
\unskip\
\newblock
\APACrefYearMonthDay{2016}{{\APACmonth{04}}}{}.
\newblock
{\BBOQ}\APACrefatitle {Recent near-{Earth} supernovae probed by global deposition of interstellar radioactive {60Fe}} {Recent near-{Earth} supernovae probed by global deposition of interstellar radioactive {60Fe}}.{\BBCQ}
\newblock
\APACjournalVolNumPages{Nature}{532}{7597}{69--72}.
\newblock
\begin{APACrefURL} [{2025-08-25}]\url{https://www.nature.com/articles/nature17196} \end{APACrefURL}
\newblock
\APACrefnote{Publisher: Nature Publishing Group}
\newblock
\begin{APACrefDOI} \doi{10.1038/nature17196} \end{APACrefDOI}
\PrintBackRefs{\CurrentBib}

\bibitem [\protect \citeauthoryear {%
{Wallner}%
\ \protect \BOthers {.}}{%
{Wallner}%
\ \protect \BOthers {.}}{%
{\protect \APACyear {2021}}%
}]{%
wallner_60fe_2021}
\APACinsertmetastar {%
wallner_60fe_2021}%
\begin{APACrefauthors}%
{Wallner}, A.%
, {Froehlich}, M\BPBI B.%
, {Hotchkis}, M\BPBI A\BPBI C.%
, {Kinoshita}, N.%
, {Paul}, M.%
, {Martschini}, M.%
\BDBL {}{Yamagata}, T.%
\end{APACrefauthors}%
\unskip\
\newblock
\APACrefYearMonthDay{2021}{{\APACmonth{05}}}{}.
\newblock
{\BBOQ}\APACrefatitle {{$^{60}$Fe and $^{244}$Pu deposited on Earth constrain the r-process yields of recent nearby supernovae}} {{$^{60}$Fe and $^{244}$Pu deposited on Earth constrain the r-process yields of recent nearby supernovae}}.{\BBCQ}
\newblock
\APACjournalVolNumPages{Science}{372}{6543}{742-745}.
\newblock
\begin{APACrefDOI} \doi{10.1126/science.aax3972} \end{APACrefDOI}
\PrintBackRefs{\CurrentBib}

\bibitem [\protect \citeauthoryear {%
Willenbring%
\ \BBA {} von Blanckenburg%
}{%
Willenbring%
\ \BBA {} von Blanckenburg%
}{%
{\protect \APACyear {2010}}%
}]{%
willenbring_long-term_2010}
\APACinsertmetastar {%
willenbring_long-term_2010}%
\begin{APACrefauthors}%
Willenbring, J\BPBI K.%
\BCBT {}\ \BBA {} von Blanckenburg, F.%
\end{APACrefauthors}%
\unskip\
\newblock
\APACrefYearMonthDay{2010}{{\APACmonth{05}}}{}.
\newblock
{\BBOQ}\APACrefatitle {Long-term stability of global erosion rates and weathering during late-{Cenozoic} cooling} {Long-term stability of global erosion rates and weathering during late-{Cenozoic} cooling}.{\BBCQ}
\newblock
\APACjournalVolNumPages{Nature}{465}{7295}{211--214}.
\newblock
\begin{APACrefURL} [{2025-09-15}]\url{https://www.nature.com/articles/nature09044} \end{APACrefURL}
\newblock
\APACrefnote{Publisher: Nature Publishing Group}
\newblock
\begin{APACrefDOI} \doi{10.1038/nature09044} \end{APACrefDOI}
\PrintBackRefs{\CurrentBib}

\bibitem [\protect \citeauthoryear {%
Zheng%
\ \protect \BOthers {.}}{%
Zheng%
\ \protect \BOthers {.}}{%
{\protect \APACyear {2024}}%
}]{%
zheng_modeling_2024}
\APACinsertmetastar {%
zheng_modeling_2024}%
\begin{APACrefauthors}%
Zheng, M.%
, Adolphi, F.%
, Ferrachat, S.%
, Mekhaldi, F.%
, Lu, Z.%
, Nilsson, A.%
\BCBL {}\ \BBA {} Lohmann, U.%
\end{APACrefauthors}%
\unskip\
\newblock
\APACrefYearMonthDay{2024}{}{}.
\newblock
{\BBOQ}\APACrefatitle {Modeling {Atmospheric} {Transport} of {Cosmogenic} {Radionuclide} {10Be} {Using} {GEOS}-{Chem} 14.1.1 and {ECHAM6}.3-{HAM2}.3: {Implications} for {Solar} and {Geomagnetic} {Reconstructions}} {Modeling {Atmospheric} {Transport} of {Cosmogenic} {Radionuclide} {10Be} {Using} {GEOS}-{Chem} 14.1.1 and {ECHAM6}.3-{HAM2}.3: {Implications} for {Solar} and {Geomagnetic} {Reconstructions}}.{\BBCQ}
\newblock
\APACjournalVolNumPages{Geophysical Research Letters}{51}{2}{e2023GL106642}.
\newblock
\begin{APACrefURL} [{2025-01-22}]\url{https://onlinelibrary.wiley.com/doi/abs/10.1029/2023GL106642} \end{APACrefURL}
\newblock
\APACrefnote{\_eprint: https://onlinelibrary.wiley.com/doi/pdf/10.1029/2023GL106642}
\newblock
\begin{APACrefDOI} \doi{10.1029/2023GL106642} \end{APACrefDOI}
\PrintBackRefs{\CurrentBib}

\end{thebibliography}

%Reference citation instructions and examples:
%
% Please use ONLY \cite and \citeA for reference citations.
% \cite for parenthetical references
% ...as shown in recent studies (Simpson et al., 2019)
% \citeA for in-text citations
% ...Simpson et al. (2019) have shown...
%
%
%...as shown by \citeA{jskilby}.
%...as shown by \citeA{lewin76}, \citeA{carson86}, \citeA{bartoldy02}, and \citeA{rinaldi03}.
%...has been shown \cite{jskilbye}.
%...has been shown \cite{lewin76,carson86,bartoldy02,rinaldi03}.
%... \cite <i.e.>[]{lewin76,carson86,bartoldy02,rinaldi03}.
%...has been shown by \cite <e.g.,>[and others]{lewin76}.
%
% apacite uses < > for prenotes and [ ] for postnotes
% DO NOT use other cite commands (e.g., \citet, \citep, \citeyear, \nocite, \citealp, etc.).
%

\end{document}